\documentclass[onecolumn,showpacs,preprintnumbers,amsmath,amssymb]{revtex4}

\usepackage{graphicx}
\usepackage{latexsym}
\usepackage{exscale}
\usepackage{amssymb}
\usepackage{amsmath}
\usepackage{dcolumn}
\usepackage{color}

\newcommand{\hw}{$\hbar \omega$}
\allowdisplaybreaks
\begin{document}

\title[Thermal vs nonthermal effects in silicon]{Thermal and nonthermal melting of silicon under femtosecond x-ray irradiation}

\author{Nikita Medvedev$^{1}$\footnote{Email: nikita.medvedev@desy.de \\
The paper is published in: Physical Review B 91 (5), 054113 (2015)
DOI: http://dx.doi.org/10.1103/PhysRevB.91.054113
},
	Zheng Li$^{1,2}$,
    and Beata Ziaja$^{1,3}$}

\address{1. Center for Free-Electron Laser Science, Deutsches Elektronen-Synchrotron DESY, Notkestrasse 85, D-22607 Hamburg, Germany}

\address{2. Department of Physics, University of Hamburg, D-20355, Hamburg, Germany}

\address{3. Institute of Nuclear Physics, Polish Academy of Sciences, Radzikowskiego 152, 31-342 Krak\'ow, Poland}

\begin{abstract}
As it is known from visible light experiments, silicon under femtosecond pulse irradiation can undergo the so-called 'nonthermal melting' if the density of electrons excited from the valence to the conduction band overcomes a certain critical value. Such ultrafast transition is induced by strong changes in the atomic potential energy surface, which trigger atomic relocation. However, heating of a material due to the electron-phonon coupling can also lead to a phase transition, called  'thermal melting'. This thermal melting can occur even if the excited-electron density is much too low to induce non-thermal effects. To study phase transitions, and in particular, the interplay of the thermal and nonthermal effects in silicon under a femtosecond x-ray irradiation, we propose their unified treatment by going beyond the Born-Oppenheimer approximation within our hybrid model based on tight binding molecular dynamics. With our extended model we identify damage thresholds for various phase transitions in irradiated silicon.  We show that electron-phonon coupling triggers the phase transition of  solid silicon into a low-density liquid  phase if the energy deposited into the sample is above $\sim0.65$ eV per atom. For the deposited doses of over $\sim0.9$ eV per atom, solid silicon undergoes a phase transition into high-density liquid phase triggered by an interplay between electron-phonon heating and nonthermal effects.  These thresholds are much lower than those predicted  with the Born-Oppenheimer approximation ($\sim2.1$ eV/atom),  and indicate a significant contribution of electron-phonon coupling to the relaxation of the laser-excited silicon. We expect that these results will stimulate dedicated experimental studies, unveiling in detail various paths of structural relaxation within laser-irradiated silicon.
\end{abstract}
\maketitle


\section{Introduction}
\label{Introduction}

Nonthermal melting is a well-established concept, known for over two decades both theoretically \cite{Stampfli1990,Stampfli1992,Silvestrelli1996,Zijlstra2013} and experimentally \cite{Shank1983,Sokolowski-Tinten2000,Rousse2001,Harb2006,Harb2008}. Thus, it can be surprising how well 'thermal' models could sometimes reproduce experimental observations of phase transitions triggered by laser pulse irradiation of solids \cite{Korfiatis2007,Medvedev2010b,Gan2011,Gan2013}. This controversy stimulated  intense discussions \cite{Gamaly2010,Gamaly2009}. It originates from the fact that the two approaches: thermal (relying on electron-phonon heating of the atomic system on an unchanged potential energy surface) and nonthermal (describing changes of interatomic potential surface via electron excitation while excluding electron-phonon coupling), are based on different assumptions and approximations, which are rarely studied together due to prohibitive computational complexity.

The nonthermal effects in solids are typically studied within the Born-Oppenheimer approximation \cite{Stampfli1990,Stampfli1992,Silvestrelli1996,Zijlstra2013}. The models treating electron-phonon coupling, such as atomistic-continuum models, TTM-MD \cite{Ivanov2003,Gan2011,Gan2013,Lipp2014a} include the non-adiabatic effects by adding an empirical electron-phonon coupling parameter. To our knowledge, there were only a few attempts to incorporate both effects for solids. For example, such attempts were made in a phenomenological manner in Refs.\  \cite{Korfiatis2007,Shokeen2013}.

It is generally believed that for a low deposited dose, thermal melting of a semiconductor or an insulator can be induced, while for a higher dose, when typically $\sim$10\% of the valence-band electrons are excited to the conduction band \cite{Sokolowski-Tinten2000,Korfiatis2007}, a nonthermal melting occurs. This is, however, not always the case: diamond is a counterexample. For diamond, at a lower deposited dose,  the nonthermal graphitization occurs  and not the thermal amorphization  \cite{Medvedev2013e,Medvedev2013f}. Thus, one cannot say {\em a priori} which mechanism dominates for a specific material, and a dedicated analysis is required for each case.

Apart from the conventional visible-light lasers, the 4$^{\bf th}$ generation light sources, the free electron lasers (FEL, such as FLASH \cite{Ackermann2007}, LCLS \cite{Emma2010}, SACLA \cite{Pile2011}, FERMI \cite{allaria2012}), emitting intense femtosecond x-ray pulses can shed new light on the problem. Since almost a decade, FELs have stimulated rapid advances in many scientific fields. Damage of semiconductors under a femtosecond x-ray irradiation starts with photoabsorption, which excites electrons from the valence band or, in contrast to a visible-light irradiation, from deep atomic shells, to high-energy states of the conduction band \cite{Medvedev2011a,Ziaja2012,Rethfeld2014}. The deep-shell holes (K- and L-shell holes for silicon) then decay via Auger processes. This is the dominant relaxation channel for light elements \cite{Keski-Rahkonen1974}. Auger decay of a hole leads to an excitation of an  electron from a higher shell to the conduction band, with the transfer of the excess energy from a relaxing deep-shell hole to the excited electron. The ejected photo- and Auger-electrons scatter further via inelastic channels (impact ionization of valence band or deep-shell electrons), or elastic channels (scattering on atoms or phonons). The impact-ionization cascading is the most important relaxation process for high-energy electrons. It typically occurs on a femtosecond timescale, and finishes when the electron loses its energy below the impact ionization threshold \cite{Medvedev2009a,Ziaja2005}. In contrast, the elastic electron-phonon scattering dominates for low-energy electrons, leading to significant electron energy losses only at longer (typically picosecond) timescales \cite{Lorazo2006,Rethfeld2002,Rethfeld2014}.
Apart from that, the transiently excited state of the electronic subsystem induces a change of the atomic potential energy surface. This can lead to nonthermal melting described above, i.e.,  in covalently bonded semiconductors, the enforced population of antibonding states within the conduction band can trigger an ultrafast rearrangement of atoms which then attempt to minimize the potential energy.

To study phase transitions in silicon triggered by a femtosecond FEL pulse, we use our recently developed hybrid model \cite{Medvedev2013e,Medvedev2013f} which traces nonequilibrium kinetics of electrons under ultrashort laser irradiation and the following rearrangement of atoms. The model relies on: (i)  a Monte Carlo description of nonequilibrium high-energy electrons, (ii) Boltzmann kinetic approach for low-energy electrons, and (iii) tight-binding molecular dynamics (TBMD) method tracing atomic trajectories, transient electronic band structure, and evolution of the interatomic potential energy surface.
The original TBMD method proposed by Jeschke {\em et al.}~\cite{Jeschke2001,Jeschke2002} was based on the Born-Oppenheimer approximation. However, to address thermal melting occurring via electron-lattice coupling,  it is necessary to extend the model beyond the Born-Oppenheimer approximation  \cite{Li2013a}. We accordingly modify our hybrid model \cite{Medvedev2013e,Medvedev2013f} by including the non-adiabatic coupling between electrons and the lattice, which is usually known as 'electron-phonon coupling'.
The proposed model calculates at each time step the respective transition rate from the matrix element for electron-atom (ion) coupling known in {\em ab-initio} femto-chemistry \cite{Tully1990,Hammes-Schiffer1994}. The rate is then included as a Boltzmann collision integral \cite{Rethfeld2002,Rethfeld2014} in the evolution equation for electron distribution. The extended model thus accounts for both thermal and nonthermal effects, allowing us to study their occurrence within one consistent theoretical framework. Note also that the same approach can be used in any {\em ab-initio} model, such as, e.g.,  density-functional-theory molecular dynamics.

\section{Model}
\label{Model_section}

\subsection{Hybrid model}
\label{hybrid}

The hybrid model addressing processes occurring in a semiconductor during its irradiation with VUV-rays or x--rays  combines various simulation methods. It was developed in \cite{Medvedev2013e} and described in detail in Ref.\ \cite{Medvedev2015}. The basic ideas are as follows. The atom dynamics is traced with the classical molecular dynamics simulation method (MD) with periodic boundary conditions\cite{Medvedev2013e,Jeschke1999,Jeschke2001}. This method requires a knowledge of the potential energy surface, which determines the forces acting on the atoms.

The potential energy surface together with the transient electronic band structure are calculated by a direct diagonalization of a tight binding (TB) Hamiltonian. The Hamiltonian is following the evolution of the atomic configuration within a simulation box, thus, changing in time.
The forces acting on atoms and the electron-lattice (electron-phonon) energy exchange depend additionally on the specific state of the electronic subsystem. The corresponding potential energy surface is calculated as in Refs. \cite{Jeschke1999,Jeschke2001}:
\begin{equation}{}
 \Phi(\{ r_{ij}(t)\}, t) = \sum_{\rm i} f_e(E_{\rm i}, t) E_{\rm i} + E_{\rm rep}( \{ r_{ij} \} )  \ ,
 \label{PotEn}
\end{equation}
where the repulsive part is describing the effective repulsion of atomic cores, $E_{rep}(\{ {\bf r}_{i j}\})$. This potential energy is used within the Parrinello-Rahman Lagrangian under the constant-pressure condition. The corresponding equations of motion are presented in \cite{Parrinello1980,Jeschke2001,Medvedev2013e}.

The transient electron distribution function $f_e(E_{\rm i}, t)$ enters Eq.~(\ref{PotEn}). Therefore, one has to trace simultaneously the evolution of the state of electronic ensemble. Additionally, the transient energy levels $E_{\rm i}$ are obtained by a diagonalization of the tight binding Hamiltonian (for more details see \cite{Medvedev2013e}).

After a VUV- or x-ray irradiation, the transient electron distribution function has a shape of the so-called 'bump-on-hot-tail'-distribution \cite{Chapman2011,Medvedev2011a,Faustlin2010,Ziaja2012,Hau-Riege2013}. This typical shape consists of nonthermalized high-energy electrons, and of (nearly-) thermalized low-energy electrons within the valence and the bottom of the conduction band. Utilizing this fact, we split the electron ensemble in two parts, treating each of them with a dedicated (computationally efficient) method.  The Monte Carlo (MC) method is used to describe the transient nonequilibrium kinetics of high-energy electrons and their secondary cascading as well as the photoabsorption and Auger decays of atomic deep-shell holes \cite{Medvedev2013e,Medvedev2009a,Ziaja2005,Ziaja2012,Medvedev2011a}. More details on the Monte Carlo model and the cross sections used can be found in \cite{Medvedev2013e, Medvedev2015}. The cross sections used for electron scattering in silicon can be found in Ref.~\cite{Rymzhanov2014a}.

A simplified Boltzmann equation is applied to describe low-energy electrons. The high-energy-electron and the low-energy-electron domains are interconnected, as electrons can gain or lose energy and go from one domain to another. This forms the source/sink terms for the low-energy part \cite{Osmani2011,Ridgway2013}, as the changing number and energy of low-energy electrons affect directly their distribution. Additionally, atomic motion and the evolution of the electronic band structure also influence the temperature of electrons and their chemical potential \cite{Medvedev2013e}.

We developed a dedicated technique to treat electron-atom energy exchange via non-adiabatic channel (electron-phonon coupling), as it will be explained in detail in section~\ref{TTM}. Such combined treatment enables us to trace modification of the atomic potential caused by the excitations of electrons and  the electron-atom energy exchange, addressing possible thermal and nonthermal phase transitions simultaneously.

\subsection{Boltzmann equation for low-energy electrons}
\label{TTM}

Evolution of the electron distribution function on the energy levels obtained from the diagonalization of the TB-Hamiltonian can be traced by means of Boltzmann collision integrals \cite{Rethfeld2010}:
\begin{eqnarray}{}
 \frac{df_i}{dt} = \sum_j I_{i,j}^{e-e} + \sum_j I_{i,j}^{e-at} + S \ ,
\label{Eq:Boltzmann}
\end{eqnarray}
where $I_{i,j}^{e-e}$ is an electron-electron collision integral, $I_{i,j}^{e-at}$ is an electron-atom collision integral, and $S$ is a source term, describing the electrons arriving and leaving the low-energy domain, in accordance with the introduced separation of low- and high-energy domains \cite{Medvedev2013e,Osmani2011,Ridgway2013,Medvedev2013}.

As the low-energy electrons are in a nearly thermalized state already after a few femtoseconds since the beginning of the laser pulse, the electron-electron collision integral turns to zero. In our model, we ensure the electron distribution function to be a Fermi-function due to the assumed instant thermalization of electrons at each time-step, similarly to  Ref. \cite{Medvedev2013e}. Note that any possible slight deviation of the exact distribution function from the equilibrium Fermi-shape affects only negligibly the atomic motion and the phase transition \cite{HaraldO.Jeschke2000}.

For the Fermi distributions one can define the corresponding chemical potential and electronic temperature \cite{Medvedev2013e}. This temperature is generally different from the atomic temperature, thus, electron-atom scattering integral governs the energy flow, $Q$,  between the two systems:
\begin{eqnarray}{}
Q = \sum_{i,j} I_{i,j}^{e-at} \cdot E_{\rm i} \ .
 \label{Eq:Heatflow}
\end{eqnarray}

The electron-atom scattering integral depends on the transient distributions of electrons and of atoms, and on the matrix element describing their interaction. Within the Born-Oppenheimer approximation, an atomic motion does not trigger any electron transition between the energy levels, as electrons are assumed  to adjuste instantly to a new configuration. Thus, the approximation is inherently incapable of reproducing electron-atom energy exchange, and the collision integral $I_{i,j}^{e-at} \equiv 0$. In order to trace electron-atom coupling, non-adiabatic effects (electron-phonon coupling) must be explicitly included. Generally, electron-atom collision integral can be written in the form (similar to \cite{Mueller2013}):
\begin{eqnarray}{}
\label{Eq:Fin_coll_int}
&&\hspace{-1em} \sum_{j=1}^{N} I^{e-at}_{i,j} = \frac{2\pi}{\hbar} \sum_{j=1}^{N} |M_{e-at}(E_i,E_j)|^2 \times \\
&&\hspace{-1em} \begin{cases}
f_e(E_i)(2 - f_e(E_j)) - f_e(E_j)(2 - f_e(E_i))g_{at}(E_i - E_j) \ ,
\nonumber\\ {\rm for} \ i > j , \\
f_e(E_i)(2 - f_e(E_j))g_{at}(E_j - E_i) - f_e(E_j)(2 - f_e(E_i)) \ ,
\nonumber\\ {\rm for} \ i < j ,\
\end{cases}
\end{eqnarray}
where $g_{at}$ is the integral of the atomic distribution function, and $M_{e-at}(E_i,E_j)$ is the matrix element for electron-atom (ion) scattering. We calculate the matrix element with the method used in non-adiabatic molecular dynamics applied in femtochemistry \cite{Tully1990,Hammes-Schiffer1994}:
\begin{eqnarray}{}
&& M_{e-at}(E_i,E_j) = \nonumber \\
&& \frac{1}{2} \left( \langle i(t-\delta t) | j(t) \rangle - \langle j(t-\delta t) | i(t) \rangle \right) (\overline{E_j} - \overline{E_i}) \ ,
 \label{Eq:Fin_matrix}
\end{eqnarray}
where $M_{e-at}(E_i,E_j)$ is the matrix element for electron transition between the levels $E_i$ and $E_j$ induced by the atomic motion, taken as a mean value on the current and previous time-steps: $\overline{E_l} = (E_l(t) + E_l(t - \delta t))/2$
; $ | i(t) \rangle$ is the electron wave-function at the time-instance $t$ obtained as an eigenfunction of the TB-Hamiltonian. Note that the electron wave-functions from two sequential time-steps ($t - \delta t$ and $t$) are entering Eq.(\ref{Eq:Fin_matrix}) \cite{Li2013a}. The derivation of the collision integral Eq.(\ref{Eq:Fin_coll_int}), the matrix element, Eq.(\ref{Eq:Fin_matrix}), and the numerical details of their calculations are presented in Appendix.

Once we calculate the average heat flow between the electrons and ions, Eq.(\ref{Eq:Heatflow}), we use the velocity scaling for atoms to accommodate the excess energy transferred from (or to) electrons on the current timestep \cite{Rapaport2004}. The corresponding change of the electron energy is introduced to the electron distribution function.

\begin{figure}
  \centering
   \includegraphics[width=0.5\textwidth]{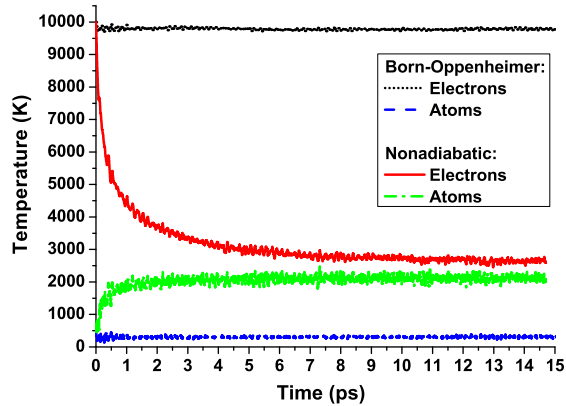}
    \caption{Electron-lattice thermalization: comparison of the non-adiabatic scheme with the Born-Oppenheimer approximation. Initial system conditions are: the electron temperature $T_e = 10000$ K, the atomic temperature $T_{at}=300$ K. A super-cell of a constant volume with 216 atoms is used.}
  \label{Pic:BO_VS_SH}
\end{figure}

A test-case is presented in Fig.\ref{Pic:BO_VS_SH}. It shows the relaxation of the electron-atom system at an initial nonequilibrium between hot electrons and room-temperature atoms.  Electron and atom temperatures are $T_e=10000$ K and $T_a=300$ K, correspondingly. The number of atoms within a unit cell is 216. It is sufficiently large so that there is no artificial influence of the number of atoms on the thermalization timescale (detailed convergence study is presented in Appendix). The Born-Oppenheimer approximation does not allow for any energy exchange between the electrons and lattice, while the non-adiabatic scheme yields reasonable timescales for electron-lattice thermalization, similar to the empirical estimates \cite{Chen2005,Gan2011,Gan2013}.


\section{Results}
\label{Results_section}

\subsection{Interplay of thermal and nonthermal effects}
\label{Sec:Thermal}

With the extended model we have simulated evolution of laser-irradiated silicon at various radiation doses absorbed per atom.

\begin{figure}[!t]
 \centering
   \includegraphics[trim=0 20 120 10,width=0.45\textwidth]{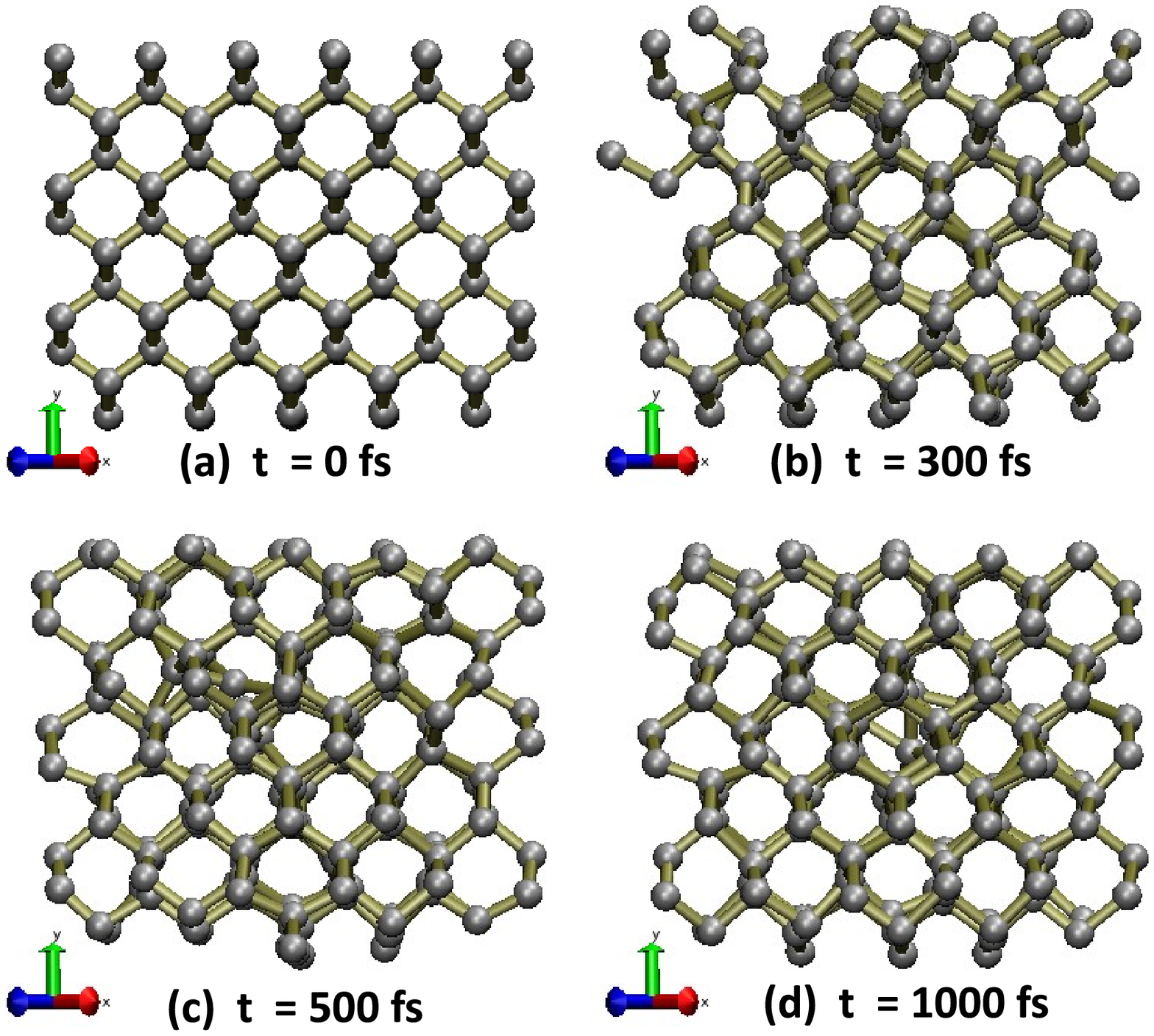}
    \caption{Transition to LDL phase: snapshots of atomic positions in silicon irradiated with 10 fs laser pulse of \hw =1 keV photon energy at the absorbed dose of 0.7 eV/atom: (a) t = 0 fs, (b) t = 300 fs, (c) t = 0.5 ps, and (d) t = 1 ps. X, Y, and Z axes are shown (left-bottom of each panel).}
  \label{Pic:0.65eVatom}
\end{figure}

\begin{figure*}[!t]
  \centering
   \includegraphics[width=0.95\textwidth]{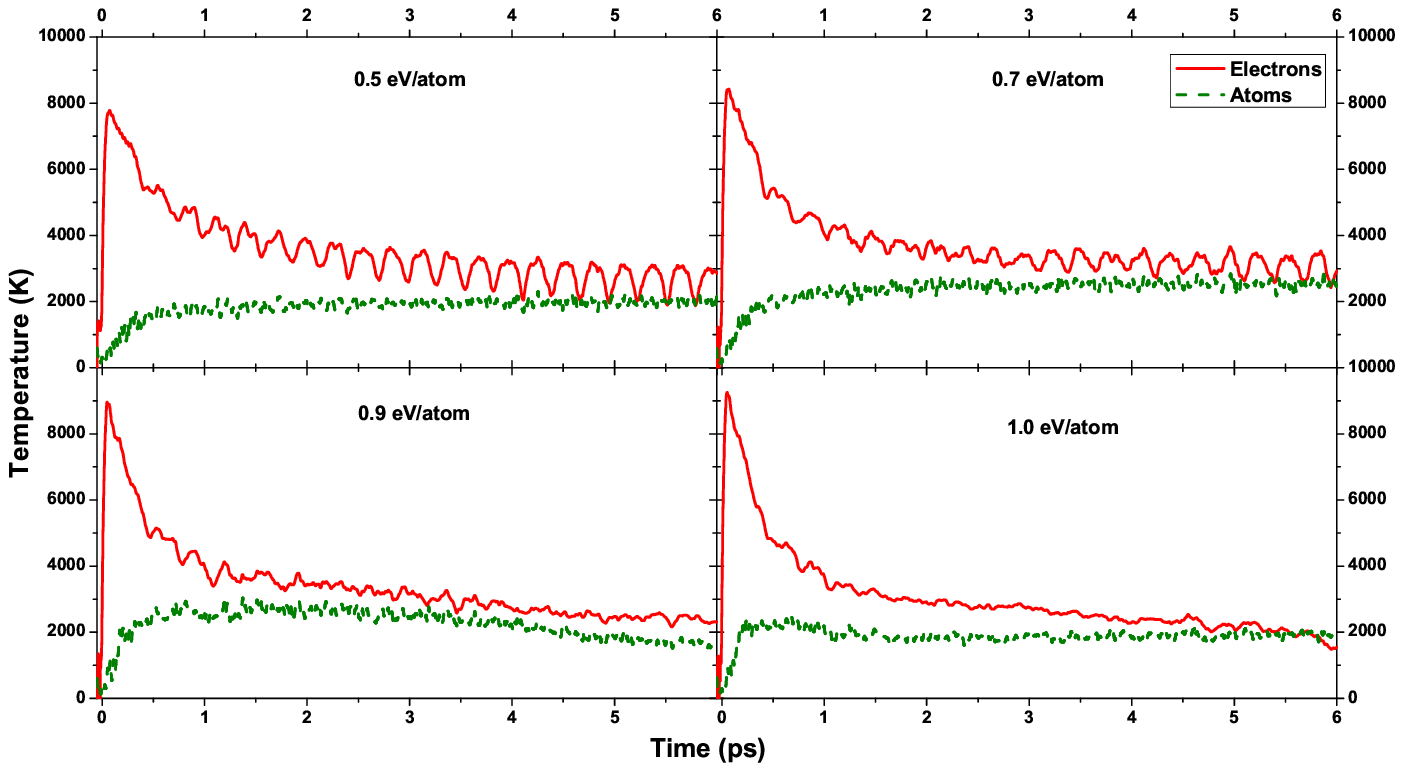}
    \caption{Electronic and atomic temperatures in silicon irradiated with 10 fs FEL pulse of  \hw=1 keV photon energy at the absorbed doses of 0.5, 0.7, 0.9 and 1 eV/atom. Silicon melting temperature is $1687$ K.}
  \label{Pic:Temperatures}
\end{figure*}

Fig.\ref{Pic:0.65eVatom} shows snapshots of the atomic positions of silicon after an FEL pulse of 10 fs duration, \hw=1 keV, and the absorbed dose of 0.7 eV/atom. At such deposited doses, silicon reaches only the low-density liquid phase (LDL) \cite{Beye2013}, characterized by an electronic phase transition into a semi-metallic state with closed band gap (below). Atomic temperature then exceeds silicon melting temperature of $1687$ K (Fig.\ref{Pic:Temperatures}). The local order in the atomic structure is preserved. Silicon remains in the LDL state during the whole simulation time, i. e.,  up to 50 ps (not shown).
\begin{figure}[!t]
  \centering
   \includegraphics[trim=0 20 120 10, width=0.45\textwidth]{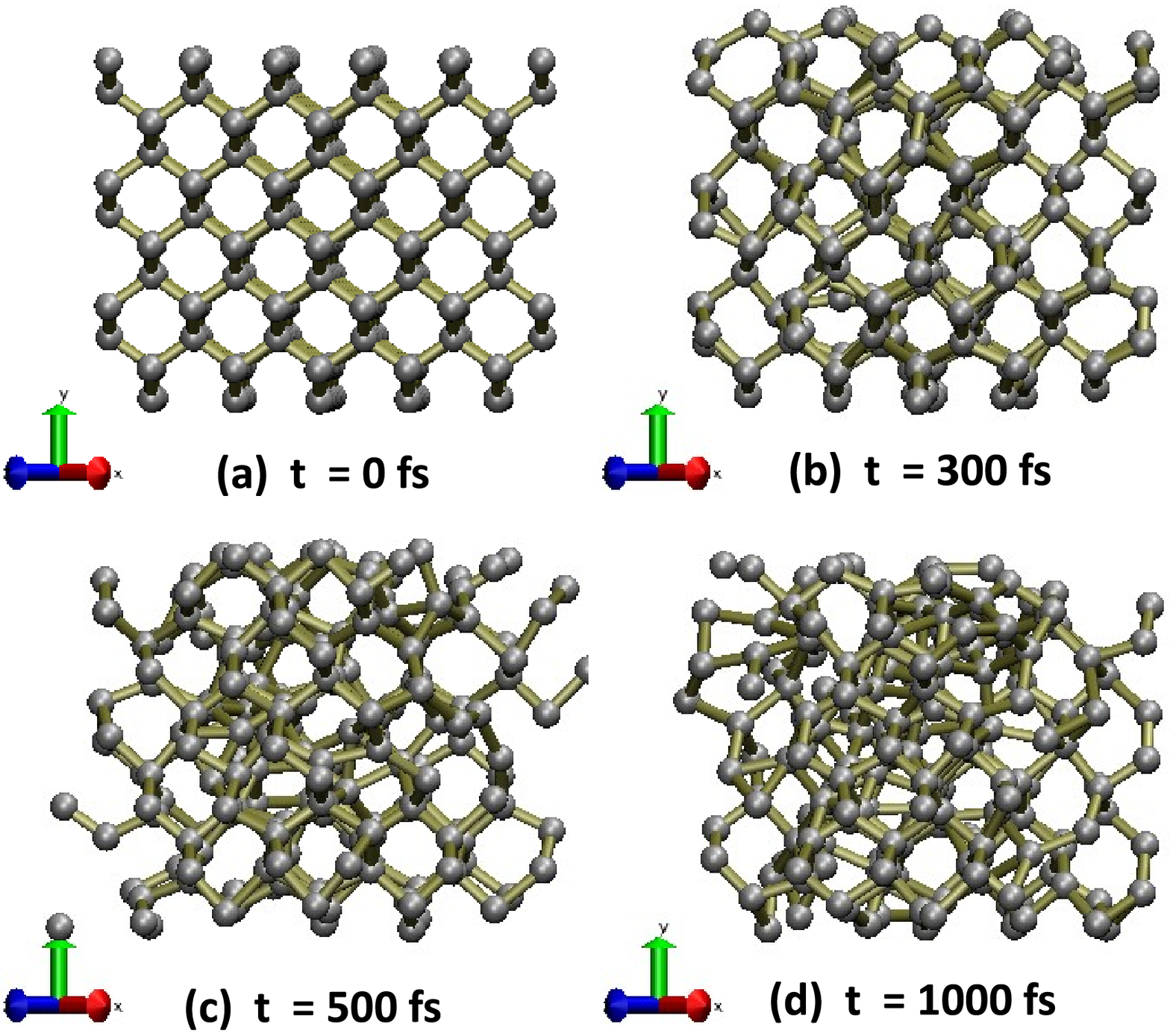}
    \caption{Transition to HDL phase: snapshots of atomic positions in silicon irradiated with 10 fs laser pulse of \hw =1 keV photon energy at the absorbed dose of 0.9 eV/atom: (a) t = 0 fs, (b) t = 300 fs, (c) t = 0.5 ps, and (d) t = 1 ps. X, Y, and Z axes are shown (left-bottom of each panel).}
  \label{Pic:0.9eVatom}
\end{figure}

The snapshots of atomic positions within silicon irradiated with an FEL pulse of 10 fs duration, \hw = 1 keV, and the absorbed  dose of 0.9 eV/atom are shown in Fig.\ref{Pic:0.9eVatom}. After absorbing this dose, silicon reaches the high-density liquid (HDL) phase with amorphization \cite{Beye2013}. This is a result of the interplay between thermal heating and nonthermal changes in the interatomic potential. They trigger atomic relocations on an short time-scales of $300-500$ fs. These time-scales match very well the experimental observations of nonthermal melting \cite{Sokolowski-Tinten2000,Harb2006,Harb2008,Sokolowski-Tinten1995}.
Similarly to observed in \cite{Beye2013}, the HDL phase is reached after an intermediate LDL phase.

\begin{figure}[!t]
  \centering
   \includegraphics[width=0.5\textwidth]{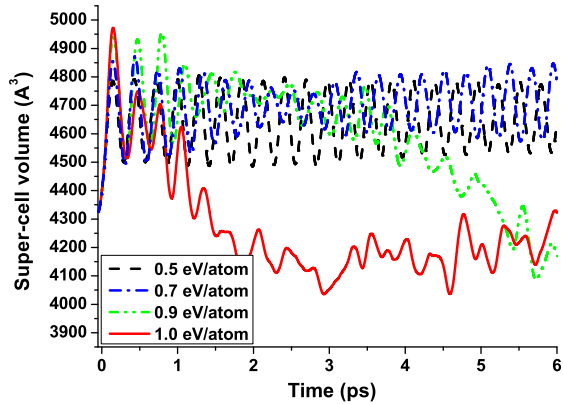}
    \caption{Volume of the super-cell of silicon (216 atoms) irradiated with 10 fs laser pulse of  \hw=1 keV photon energy at various absorbed doses.}
  \label{Pic:Volumes}
\end{figure}

\begin{figure}[!t]
  \centering
   \includegraphics[width=0.5\textwidth]{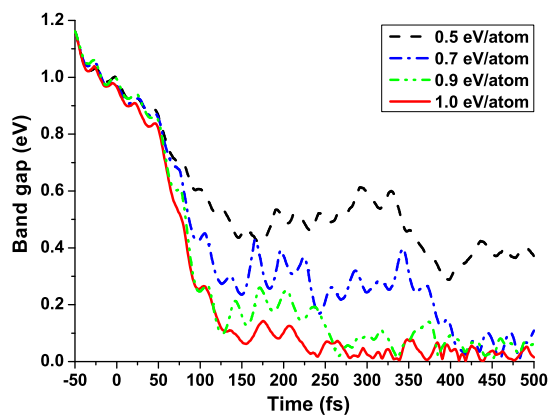}
    \caption{Band gap of silicon irradiated with 10 fs laser pulse of  \hw=1 keV photon energy  plotted as a function of time at various absorbed doses.}
  \label{Pic:Band_gaps}
\end{figure}

Figure~\ref{Pic:Volumes} shows how the volume of the Parrinello-Rahman super-cell changes after the irradiation with different fluences. As the number of atoms and ions within the super-cell is conserved, the decrease or increase of the super-cell volume corresponds to the increase or decrease of the atomic density respectively. For the doses of 0.5 and 0.7 eV  per atom which are below the nonthermal damage threshold, one can see an increase of the super-cell volume, corresponding to the decrease of  the atomic density only during the electron cascading time ($\sim$first hundreds of fs). The dose of 0.7 eV/atom corresponds to the phase transition to the LDL phase, as  indicated by the band gap collapse in Fig.\ref{Pic:Band_gaps}. The doses above the nonthermal melting threshold, 0.9 and 1 eV/atom in Fig.\ref{Pic:Volumes}, induce a transient increase of the volume with its shrinkage later. This reflects the intermediate transition of silicon to the LDL phase followed by the transition to HDL phase.

Fig.~\ref{Pic:Band_gaps} presents the band gap of silicon after different energy depositions. Band gap is defined as the energy difference between the closest eigenstates above and below the Fermi energy.
Band gap width shrinks to nearly zero already at $\sim 0.7$ eV/atom, showing that silicon is in a semi-metallic state (LDL). This is in agreement with our results indicating that the phase transition into LDL phase occurs above the threshold of $\sim 0.65$ eV/atom (Fig.~\ref{Pic:0.65eVatom}).

\begin{figure}[!t]
  \centering
   \includegraphics[width=0.5\textwidth]{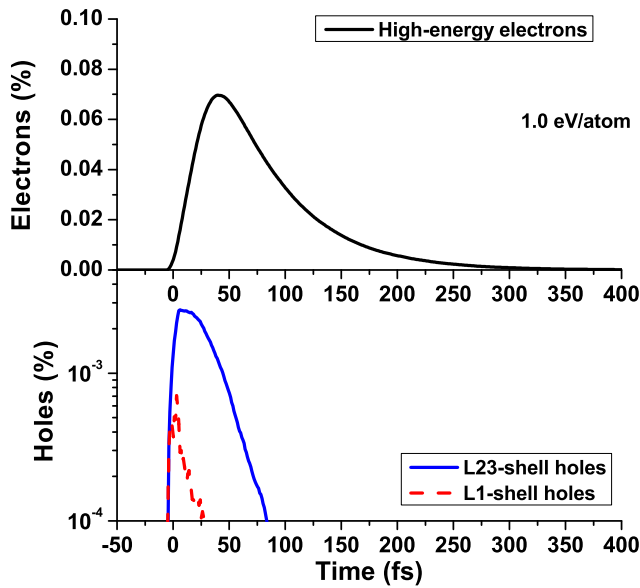}
    \caption{Densities of high-energy electrons (top panel) and holes in $L_1$ and $L_{2,3}$-shell holes (bottom-panel)  as a function of time. They are expressed as a percentage of the initial valence-electron density.}
  \label{Pic:HE_electrons}
\end{figure}

\begin{figure*}[!t]
  \centering
   \includegraphics[width=0.95\textwidth]{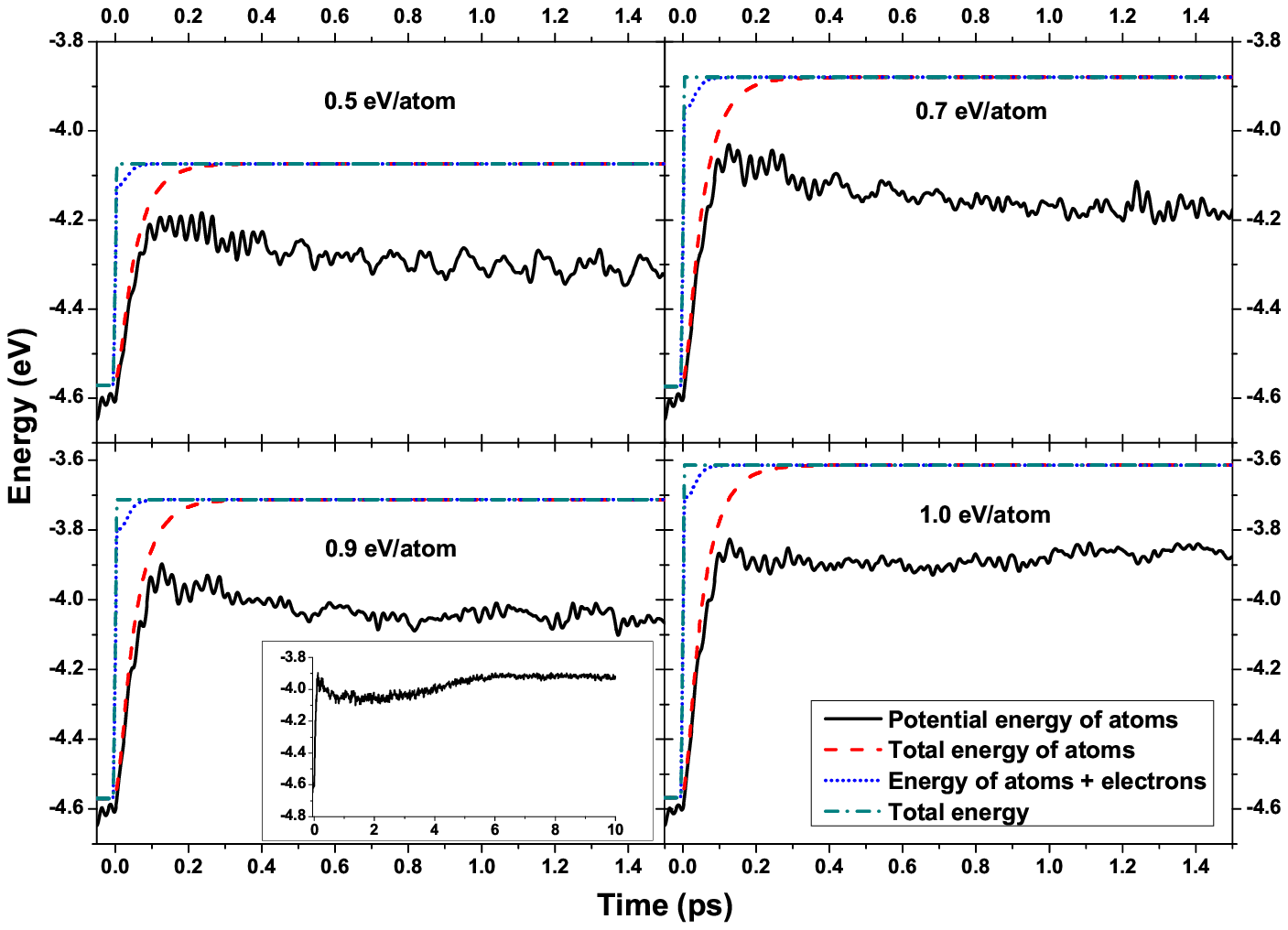}
    \caption{Energy redistribution in silicon irradiated with 10 fs FEL pulse of  \hw=1 keV photon energy at the absorbed doses of 0.5, 0.7, 0.9 and 1 eV/atom. The black solid line is the potential energy of atoms; red dashed line is the total energy of atoms (potential and kinetic); blue dotted line is the total energy of atoms and electrons (energy of the system excluding deep-shell holes); and the green dash-dotted line is the total energy of the system. The inset is showing longer timescales behavior of the potential energy of atoms for the case of 0.9 eV/atom absorbed dose.}
  \label{Pic:Energies}
\end{figure*}

Fig.~\ref{Pic:HE_electrons} shows that deep-shell holes decay quickly via Auger-decay, as discussed above, and their energy is brought back to the electronic system on a sub-100 fs scale.  Electron cascading finishes within $\sim 250-300$ fs (also seen in Fig.~\ref{Pic:HE_electrons}), and the energy is transferred to the low-energy electrons. The figure~\ref{Pic:Energies} confirms that the energy redistribution between different subsystems of the irradiated material occurs on femtosecond scale, starting with the excitation of electrons and holes. The total energy is conserved, as we did not consider any energy transport from the system, assuming periodic boundaries.
In contrast to the diamond-to-graphite phase transition reported in \cite{Medvedev2013e,Medvedev2013f}, the atomic potential energy does not exhibit a rapid jump at the beginning of phase transition. Instead, the system is relaxing on picosecond timescales. For higher absorbed doses (0.9 and 1 eV/atom in figure~\ref{Pic:Energies}), silicon turns into the high-density liquid phase, which is also reflected by the potential energy curve: it is slightly raising (inset in the Fig.~\ref{Pic:Energies}), while atomic temperature is decreasing (see below in Fig.~\ref{Pic:Temperatures}). The timescales of these changes match the timescales of the super-cell volume contraction shown in Fig.~\ref{Pic:Volumes}.

Atomic and electronic temperatures for the same absorbed doses are shown in Fig.~\ref{Pic:Temperatures}. The electron temperature is increasing during the laser pulse and during the relaxation of high-energy electrons and deep-shell holes (first $200$ fs). Later, while electrons transfer their energy to the lattice, the electron temperature is decreasing. This takes a few picoseconds. After that time, electrons are equilibrated with the atoms. The electron temperature oscillations (especially pronounced for low dose irradiation) are caused by the super-cell volume oscillations, hence, they can be considered as a simulation artifact. Such oscillations are reflected either in electron temperature (for fixed energy), or in the energy (for fixed temperature); we chose the first scheme in the simulation. Atomic temperature oscillations reflect physical process: the exchange between kinetic and potential energies of atoms.
At the highest fluence, atomic temperature increases rapidly already within $\sim300$ fs. This reflects a strong interplay between thermal and nonthermal effects within the system.

\begin{figure}[!t]
  \centering
   \includegraphics[width=0.5\textwidth]{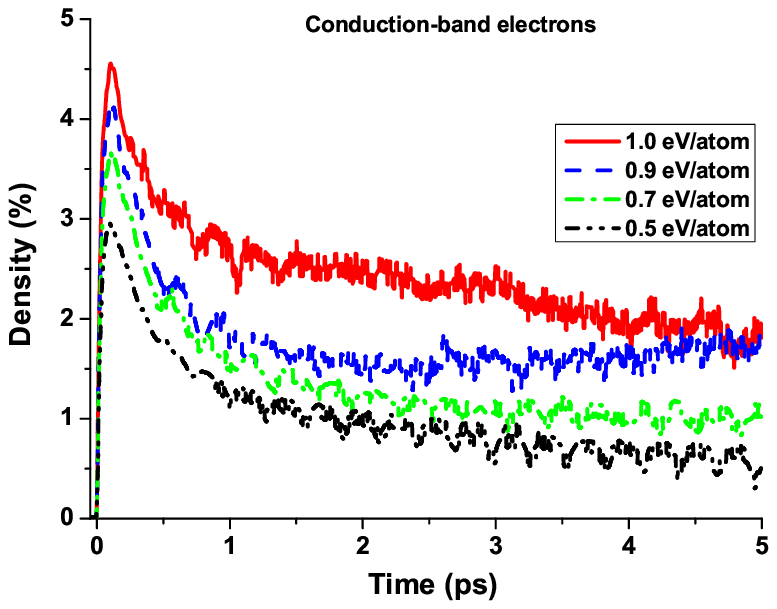}
    \caption{Density of low-energy conduction-band electrons in silicon after an FEL pulse of 10 fs duration,  \hw=1 keV photon energy at the absorbed doses of 0.5, 0.7, 0.9 and 1 eV/atom.}
  \label{Pic:CB_electrons}
\end{figure}

The number of low-energy conduction-band electrons is shown in Fig.~\ref{Pic:CB_electrons}. For the absorbed doses considered above it never reaches the critical value of $9 \%$ which purely nonthermal models predict  \cite{Stampfli1990,Stampfli1992}. However, the number of electrons  is sufficiently high to trigger a phase transition, which is then due to the thermal heating of the system with nonthermally-weakened interatomic bonds. To mention, the peak conduction-band electron densities are close to the ones estimated experimentally \cite{Harb2008}.

Note that the density of the excited electrons reached after FEL irradiation at the pulse fluences considered here, which provide absorbed dose on the level of $\sim1$ eV/atom, is of the order of a few percent of the solid density ($10^{21}-10^{22}$ cm$^{-3}$). This is a high density for an electron plasma. Its (partial) thermalization is then known to be very rapid \cite{Chapman2011,Medvedev2011a,Faustlin2010,Ziaja2012,Hau-Riege2013}, confirming the assumptions made above. On the other hand, being only a few percent of the solid density, it ensures that the applicability condition of the tight binding scheme (low excitation regime) is not violated.

\subsection{Damage threshold for silicon as a function of photon energy}
\label{Sec:Damage}

\begin{figure}[!t]
  \centering
   \includegraphics[width=0.5\textwidth]{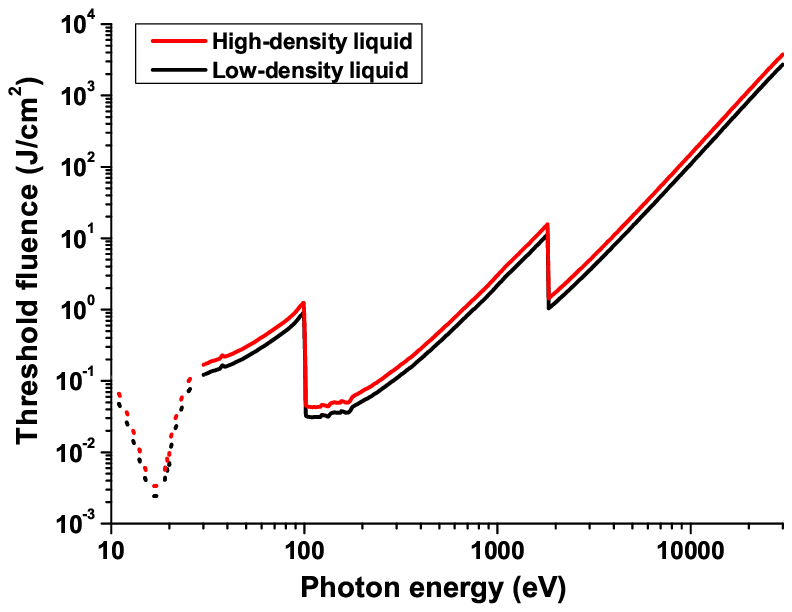}
    \caption{Damage threshold fluences for silicon corresponding to the low-density liquid and high-density liquid formation as a function of photon energy.}
  \label{Pic:Threshold}
\end{figure}

Following the procedure from Ref. \cite{Medvedev2013f}, we estimate the damage threshold of silicon for different photon energies. We checked (not shown) that the damage threshold in terms of deposited energy per atom is almost independent of the incoming photon energy. Thus, by converting the dose per atom into the units of incoming fluence with one-photon absorption cross section from Refs. \cite{Palik1985,Henke1993,Cullen1997}, we obtain the damage threshold predictions shown in Fig.~\ref{Pic:Threshold}. They can be directly verified experimentally.
No effects of thermal diffusion and particle diffusion were taken into account for the predictions from Fig.~\ref{Pic:Threshold}. They can play a role for the case of small skin-depth (the minimum around 20 eV on the picture). Due to these effects, and possible re-solidification governed by the energy flows out of the laser spot, at such photon energies our calculations might underestimate the experimentally measured damage thresholds.

\subsection{Purely nonthermal melting of silicon}
\label{Sec:Nonthermal}

Finally, for comparison we show the predictions for silicon melting obtained after excluding the non-adiabatic effects (electron-phonon coupling). Within the adiabatic (Born-Oppenheimer) scheme the calculated damage threshold for nonthermal melting of silicon appears to be at the absorbed dose of 2.1 eV/atom. This corresponds to 9\% of electrons excited from the valence to the antibonding states of the conduction band. This threshold value of the electron density is in an excellent agreement with earlier works \cite{Stampfli1990,Stampfli1992}, based on the adiabatic scheme.

\begin{figure}[!t]
  \centering
   \includegraphics[trim=0 20 80 10,width=0.45\textwidth]{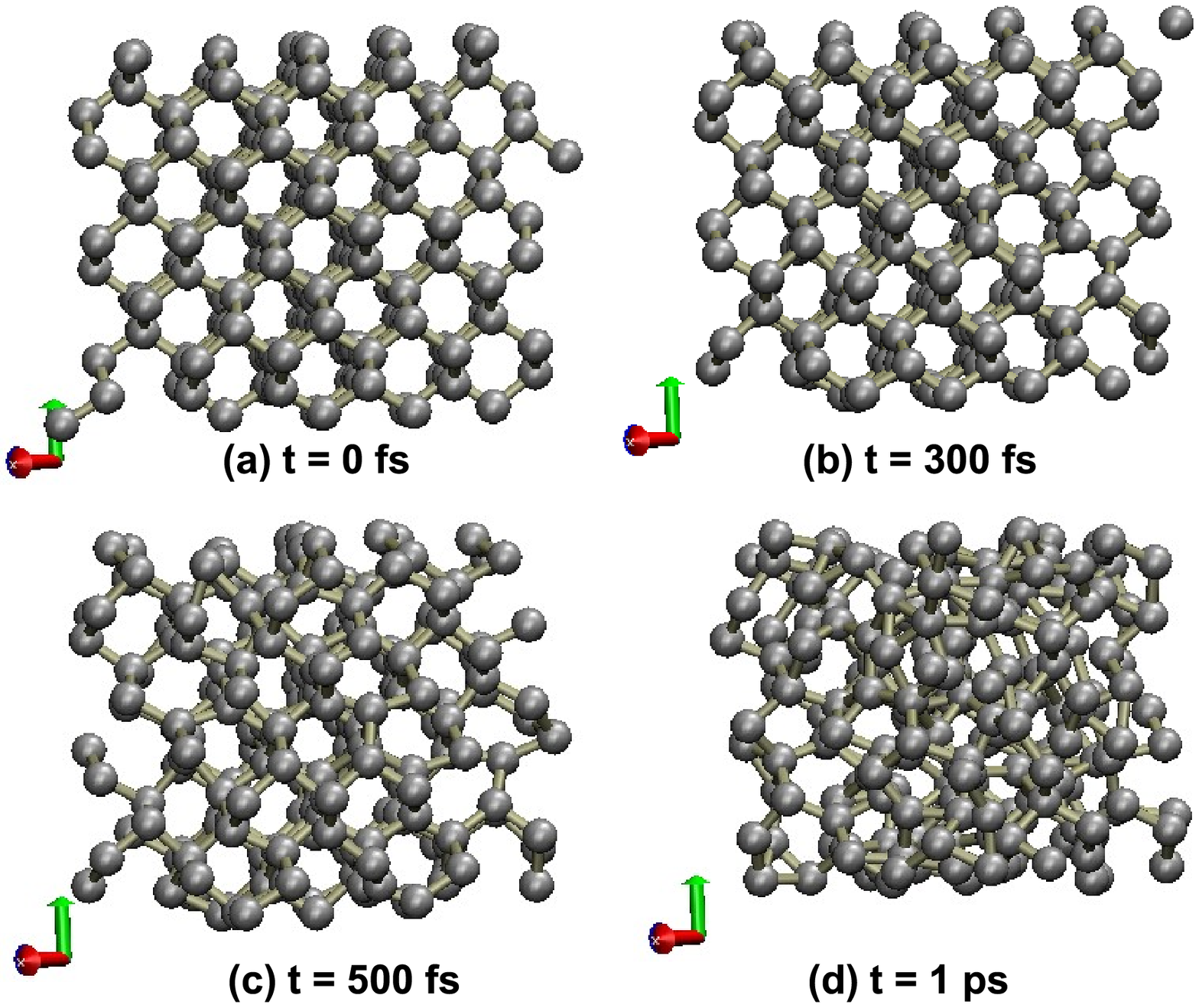}
    \caption{Nonthermal phase transition: snapshots of atomic positions in silicon irradiated with 10 fs laser pulse of  \hw = 1 keV photon energy at the absorbed dose of 2.5 eV/atom: (a) t = 0 fs, (b) t = 300 fs, (c) t=0.5 ps, and (d) t = 1 ps. X, Y, and Z axes are shown (left-bottom of each panel).}
  \label{Pic:Nonthermal}
\end{figure}

The atomic snapshots shown in Fig.~\ref{Pic:Nonthermal} demonstrate the final state to be an amorphous high-density liquid \cite{Beye2010}. The transition to that state proceeds on sub-picosecond timescales at which thermal effects - if included - would play a role. The intermediate LDL phase  can be identified by the collapsed band gap and increased volume of the modeled super-cell (decreased density) prior to the final contraction to the HDL phase.

As discussed above, the model of nonthermal melting of silicon based on the Born-Oppenheimer approximation does not allow for electron-phonon coupling, i.e.,  the excited electrons do not exchange energy with atoms.  The damage then occurs as a purely nonthermal effect of weakening the interatomic bonds due to electron excitation. This results in much higher damage threshold than that obtained with non-adiabatic approach. These finding also indicates that thermal models such as \cite{Medvedev2010b,Gan2011,Gan2013,Lipp2014a} can be accurately
applied for silicon irradiated with laser pulses of low  fluences,  when the nonthermal effects do not play a significant role.

\section{Conclusion}
\label{Conclusions_section}

In the present work, we studied phase transitions in silicon under a femtosecond irradiation with x-ray laser. In order to account both for thermally and nonthermally triggered transitions, we have extended our recently developed hybrid model  \cite{Medvedev2013e,Medvedev2013f} by including non-adiabatic electron-phonon coupling. In this way heating of a material due to the electron-phonon coupling  can also  be treated. The developed scheme is general and can be used in any {\em ab-initio} molecular dynamics model.

We demonstrated that for silicon under a femtosecond x-ray irradiation, the non-adiabatic energy exchange triggers a phase transition into low-density liquid phase above the threshold of $\sim 0.65$ eV per atom in terms of the absorbed dose. This semi-metallic state is characterized by a closed band gap, with the local order present in atomic structure. At higher doses above $\sim 0.9$ eV/atom, silicon melts into high-density liquid phase with amorphous atomic arrangement. The modeled phase transition occurs within $\sim 300-500$ fs, in a good agreement with the timescales observed in experiments. We have also predicted the damage threshold fluence in silicon as a function of the incoming photon energy.

The transition into high-density liquid phase proceeds as a result of the interplay between nonthermal and thermal effects. Weakening of interatomic bonds and heating of lattice by excited electronic subsystem triggers ultrafast amorphization of silicon. Neglecting electron-phonon coupling results in a significant overestimation of the phase transition threshold, which then is $\sim 2.1$ eV/atom. This threshold discrepancy indicates that the non-adiabatic effect has to be taken into account in the description of  the transitions within x-ray excited silicon.  Future experiments with FEL should be able to verify these predictions and unveil more details on the radiation-triggered structural transitions in silicon.


\section{Acknowledgments}
The authors thank H. O. Jeschke, V. Lipp, R. Santra, R. Sobierajski, O. Vendrell and W. Wurth for illuminating discussions.

\appendix 
\setcounter{section}{1}

\section*{Appendix: Non-adiabatic electron-atom energy exchange}

Boltzmann electron-atom collision integral, $I_{e-at}$,  can be generally written in the following form (similar to \cite{Mueller2013}):
\begin{eqnarray}{}
 \label{Coll_int}
\sum_{j=1}^{N} I^{e-at}_{i,j} = \frac{2\pi}{\hbar} \sum_{j=1}^{N} \int_{0}^{\infty} \int_{0}^{\infty} |M_{e-at}(E_i,E_j)|^2
\ F_{i,j}(E_i,E_j,E_{at},E_{at}^{fin}) dE_{at} dE_{at}^{fin}, \
\end{eqnarray}
where
$$ F_{i,j} =
\begin{cases}
(f_e(E_i)(2 - f_e(E_j))f_{at}(E_{at}) - f_e(E_j)(2 - f_e(E_i))f_{at}(E_{at}^{fin})) 
\delta(E_{at}^{fin} - E_{at} + E_i - E_j) \ , {\rm for} \ i > j \\
(f_e(E_i)(2 - f_e(E_j))f_{at}(E_{at}^{fin}) - f_e(E_j)(2 - f_e(E_i))f_{at}(E_{at})) 
\delta(E_{at}^{fin} - E_{at} + E_j - E_i)
\ , {\rm for} \ i < j ,\
\end{cases}
$$
is split up into two intervals depending on indices $i$ and $j$, and summations are running through all the $N$ energy states. These state in our case are obtained by diagonalization of the tight binding Hamiltonian. $M_{e-at}$ is the electron-atom scattering matrix element; $f_e(E)$ is the electron distribution function, assumed to be here the Fermi function, normalized to 2 accounting for the electron spin; $f_{at}$ is the atom distribution function, taken at the initial, $E_{at}$, and final, $E_{at}^{fin}$, states.

After the integration over $E_{at}^{fin}$, the energy conservation condition gives the following:
\begin{eqnarray}
&& \int_{0}^{\infty} F_{i,j}(E_i,E_j,E_{at},E_{at}^{fin}) dE_{at}^{fin} = \\
&& \begin{cases}
f_e(E_i)(2 - f_e(E_j))f_{at}(E_{at}) -
 \  f_e(E_j)(2 - f_e(E_i))f_{at}(E_{at} - (E_i - E_j)) \theta(E_{at} - (E_i - E_j)) , \ {\rm for} \ i > j \\
f_e(E_j)(2 - f_e(E_i))f_{at}(E_{at} - (E_j - E_i)) \theta(E_{at} - (E_j - E_i))  -
 \  f_e(E_i)(2 - f_e(E_j))f_{at}(E_{at}) , \ {\rm for} \ i < j ,\
\nonumber
\end{cases}
 \label{Coll_int_split}
\end{eqnarray}
where the $\theta$-functions are introduced ($\theta(x) = 0$ for $x<0$, and $\theta(x)=1$ otherwise). These $\theta$-functions are allowing only for the transitions that conserve energy: in case when an electron jumps up to the final energy level above the initial one, $j>i$, the transitions are only possible as long as the atomic energy, $E_{at}$, is sufficient to contribute to such an event.

For numerical evaluation of the collision integral over $E_{at}$ (Eq.(\ref{Coll_int})) we assume that the atomic distribution is the Maxwellian one with the transient temperature calculated as a kinetic temperature of atoms in a box with periodic boundaries:
\begin{eqnarray}{}
&& f_{at}(E_{at}) = 2 \sqrt{\frac{E_{at}}{\pi}}\cdot \frac{1}{T_{at}^{3/2}}\cdot \exp{\left(-\frac{E_{at}}{T_{at}}\right)} \ , \\
&& T_{at} = \frac{2 E_{kin}}{3N_{at} - 6} \ ,\
 \label{Coll_int_split2}
\end{eqnarray}
where $E_{kin}$ is the total kinetic energy of atoms in the super-cell; $N_{at}$ is the number of atoms; $T_{at}$ is the atomic temperature in energy units. This assumption is justified as the atomic heating is a slow process compared to nonthermal melting, which does not bring the system far out of thermal equilibrium. The initially thermalized atomic system only gains kinetic energy, thus increasing the temperature, but it remains in the equilibrium state described by the Maxwellian distribution.

Although one could, in principle, obtain the transient nonequilibrium distribution of atoms by sorting the atomic energies into energy intervals, such an approach shows itself as numerically challenging, because an integral and a double summation have to be performed in Eq.(\ref{Coll_int}). In contrast, Maxwellian distribution function for atoms allows for exact analytical evaluation of the inner integral. This also significantly speeds up the calculations.

An integral of the Maxwellian function with the $\theta$-function is as follows:
\begin{eqnarray}{}
&&  g_{at}(E) = \int_{E}^{\infty} f_{at}(E_{at}) d E_{at} =
2 \sqrt{\frac{E}{\pi T_{at}}}\cdot \exp{\left(-\frac{E}{T_{at}}\right)} - \left(erf\left(\sqrt{\frac{E}{T_{at}}}\right) - 1 \right)\ ,
\label{Maxwell_int}
\end{eqnarray}
where $erf(x)$ is the error-function. Note that an integral with the lower limit $E=0$ yields $1$.

Combining these equations together,  the total collision integral can be written as:
\begin{eqnarray}{}
\sum_{j=1}^{N} I^{e-at}_{i,j} = \frac{2\pi}{\hbar} \sum_{j=1}^{N} |M_{e-at}(E_i,E_j)|^2 
\begin{cases}
f_e(E_i)(2 - f_e(E_j)) - f_e(E_j)(2 - f_e(E_i))g_{at}(E_i - E_j) \ , {\rm for} \ i > j , \\
f_e(E_i)(2 - f_e(E_j))g_{at}(E_j - E_i) - f_e(E_j)(2 - f_e(E_i)) \ , {\rm for} \ i < j ,\
\end{cases}
\label{Fin_coll_int}
\end{eqnarray}
where the integrated Maxwellian function, $g_{at}(E)$, is defined by Eq.(\ref{Maxwell_int}).

\begin{figure}[!t]
  \centering
   \includegraphics[width=0.6\textwidth]{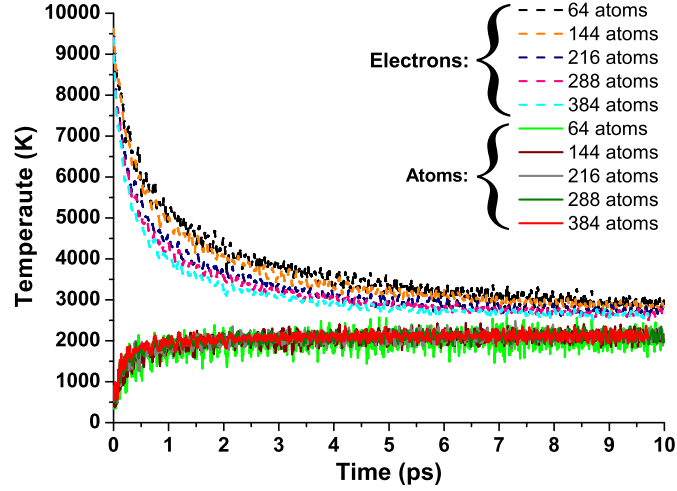}
    \caption{Convergence study of the electron-atom energy exchange rate with respect to the number of atoms in the simulation box. Initial conditions applied are: the electron temperature $T_e = 10000$ K, the atomic temperature $T_{at}=300$ K. Constant-volume simulations are used here.}
  \label{Pic:Converge}
\end{figure}

The last remaining term to be defined is the electron-atom scattering matrix element $M_{e-at}(E_i,E_j)$. To derive it, we use an {\em ab-initio} approach proposed by Tully \cite{Tully1990,Hammes-Schiffer1994}. 
Probabilities of non-adiabatic transitions of electrons between the energy levels $i$ and $j$ induced by the atomic motion during the current time step, corresponding to the atomic displacement $ \vec{R} = \vec{R}_0 + \delta \vec{R}$, can be written in the diabatic representation as follows \cite{Plasser2012}:
\begin{equation}{}
M_{e-at}(E_i,E_j) = \langle i (R_0(t-\delta t)) | \hat{H}_{e-at}(R(t)) | j (R_0(t-\delta t)) \rangle , \
 \label{Matrix_el}
\end{equation}
under an assumption of an infinitesimal time-step $\delta t$, the Hamiltonian can be expanded as $\hat{H}_{e-at}(R(t)) = \hat{H}_{e-at}(R_0(t-\delta t)) + \nabla \hat{H}_{e-at}(R_0(t-\delta t)) \cdot \delta \vec{R} $:
\begin{equation}{}
M_{e-at}(E_i,E_j) = \langle i (R_0(t-\delta t)) | \nabla H_{e-at} (R_0(t-\delta t)) | j (R_0(t-\delta t)) \rangle \ \cdot \delta \vec{R} , \
 \label{Matrix_el_2}
\end{equation}

Utilizing the Hellmann-Feynman theorem, one can rewrite the expression for the potential in terms of
the energy levels (eigenvalues of the Hamiltonian)
and
the derivatives of the wave-functions with respect to the nuclear coordinates:
\begin{equation}{}
\langle i | \nabla H_{e-at} | j \rangle = (E_j - E_i) \langle i | \nabla | j \rangle \ . \
 \label{Chain_rule}
\end{equation}

Thus, for the infinitesimal time-step $\delta t$, the matrix element can be expressed in terms of the non-adiabatic coupling vector ($\vec{d}_{i,j} = \langle i | \nabla | j \rangle$) \cite{Hammes-Schiffer1994}:
\begin{eqnarray}{}
\langle i | \nabla | j \rangle \cdot \delta \vec{R}  =\delta t \vec{\dot{R}}  \cdot \vec{d}_{i,j}
= \delta t  \frac{(\langle i(t-\delta t) | j(t) \rangle - \langle j(t-\delta t) | i(t) \rangle )}{2 \delta t}
 \label{Nabla}
\end{eqnarray}
where $\vec{\dot{R}}$ is the atomic velocity; and the wave-functions are defined at the current and the previous step of the simulation \cite{Li2013a}. The final expression for the matrix element of the non-adiabatic electron coupling to the atoms can now be written as:
\begin{eqnarray}{}
M_{e-at}(E_i,E_j) = \frac{1}{2} \left( \langle i(t-\delta t) | j(t) \rangle - \langle j(t-\delta t) | i(t) \rangle \right) (\overline{E_j} - \overline{E_i})
 \label{Fin_matrix}
\end{eqnarray}
where an energy level is taken as a mean value on the current and previous time-steps: $\overline{E_l} = (E_l(t) + E_l(t - \delta t))/2$.
In our case $\hat{H}_{e-at}(R(t))$ is equal to the tight-binding Hamiltonian, thus, the energy levels are the eigenvectors corresponding to the eigenfunctions of the tight binding Hamiltonian.
These equations (\ref{Fin_matrix}) and (\ref{Fin_coll_int}) constitute the non-adiabatic coupling between the electronic and atomic subsystems which has been used in the current work.

The finite-size effects must be investigated prior to any application of the model to a realistic situation. For this purpose, we analyzed an electron-atom energy exchange rates in a nonequilibrium model system at the following initial conditions: the electron temperature of $T_e = 10000$ K, and the atomic temperature of $T_{at}=300$ K. The results are shown in Fig.~\ref{Pic:Converge}. The figure shows that the number of atoms only slightly affects the electronic temperature, and has almost no influence on the atomic temperature. Slight decrease of the electron-atom energy exchange rate for low numbers of atoms can be attributed to the contribution of long-wavelength phonons, which appear only for sufficiently large simulation boxes. Their contribution is, however, only minor: any differences practically vanish for the number of atoms exceeding 216.


\bibliography{My_Collection}

\begin{thebibliography}{59}
\expandafter\ifx\csname natexlab\endcsname\relax\def\natexlab#1{#1}\fi
\expandafter\ifx\csname bibnamefont\endcsname\relax
  \def\bibnamefont#1{#1}\fi
\expandafter\ifx\csname bibfnamefont\endcsname\relax
  \def\bibfnamefont#1{#1}\fi
\expandafter\ifx\csname citenamefont\endcsname\relax
  \def\citenamefont#1{#1}\fi
\expandafter\ifx\csname url\endcsname\relax
  \def\url#1{\texttt{#1}}\fi
\expandafter\ifx\csname urlprefix\endcsname\relax\def\urlprefix{URL }\fi
\providecommand{\bibinfo}[2]{#2}
\providecommand{\eprint}[2][]{\url{#2}}

\bibitem[{\citenamefont{Stampfli and Bennemann}(1990)}]{Stampfli1990}
\bibinfo{author}{\bibfnamefont{P.}~\bibnamefont{Stampfli}} \bibnamefont{and}
  \bibinfo{author}{\bibfnamefont{K.}~\bibnamefont{Bennemann}},
  \bibinfo{journal}{Physical Review B} \textbf{\bibinfo{volume}{42}},
  \bibinfo{pages}{7163} (\bibinfo{year}{1990}), ISSN \bibinfo{issn}{0163-1829},
  \urlprefix\url{http://link.aps.org/doi/10.1103/PhysRevB.42.7163}.

\bibitem[{\citenamefont{Stampfli and Bennemann}(1992)}]{Stampfli1992}
\bibinfo{author}{\bibfnamefont{P.}~\bibnamefont{Stampfli}} \bibnamefont{and}
  \bibinfo{author}{\bibfnamefont{K.}~\bibnamefont{Bennemann}},
  \bibinfo{journal}{Physical Review B} \textbf{\bibinfo{volume}{46}},
  \bibinfo{pages}{10686} (\bibinfo{year}{1992}), ISSN
  \bibinfo{issn}{0163-1829},
  \urlprefix\url{http://link.aps.org/doi/10.1103/PhysRevB.46.10686}.

\bibitem[{\citenamefont{Silvestrelli et~al.}(1996)\citenamefont{Silvestrelli,
  Alavi, Parrinello, and Frenkel}}]{Silvestrelli1996}
\bibinfo{author}{\bibfnamefont{P.}~\bibnamefont{Silvestrelli}},
  \bibinfo{author}{\bibfnamefont{A.}~\bibnamefont{Alavi}},
  \bibinfo{author}{\bibfnamefont{M.}~\bibnamefont{Parrinello}},
  \bibnamefont{and} \bibinfo{author}{\bibfnamefont{D.}~\bibnamefont{Frenkel}},
  \bibinfo{journal}{Physical Review Letters} \textbf{\bibinfo{volume}{77}},
  \bibinfo{pages}{3149} (\bibinfo{year}{1996}), ISSN \bibinfo{issn}{0031-9007},
  \urlprefix\url{http://link.aps.org/doi/10.1103/PhysRevLett.77.3149}.

\bibitem[{\citenamefont{Zijlstra et~al.}(2013)\citenamefont{Zijlstra, Kalitsov,
  Zier, and Garcia}}]{Zijlstra2013}
\bibinfo{author}{\bibfnamefont{E.}~\bibnamefont{Zijlstra}},
  \bibinfo{author}{\bibfnamefont{A.}~\bibnamefont{Kalitsov}},
  \bibinfo{author}{\bibfnamefont{T.}~\bibnamefont{Zier}}, \bibnamefont{and}
  \bibinfo{author}{\bibfnamefont{M.}~\bibnamefont{Garcia}},
  \bibinfo{journal}{Physical Review X} \textbf{\bibinfo{volume}{3}},
  \bibinfo{pages}{011005} (\bibinfo{year}{2013}), ISSN
  \bibinfo{issn}{2160-3308},
  \urlprefix\url{http://link.aps.org/doi/10.1103/PhysRevX.3.011005}.

\bibitem[{\citenamefont{Shank et~al.}(1983)\citenamefont{Shank, Yen, and
  Hirlimann}}]{Shank1983}
\bibinfo{author}{\bibfnamefont{C.}~\bibnamefont{Shank}},
  \bibinfo{author}{\bibfnamefont{R.}~\bibnamefont{Yen}}, \bibnamefont{and}
  \bibinfo{author}{\bibfnamefont{C.}~\bibnamefont{Hirlimann}},
  \bibinfo{journal}{Physical Review Letters} \textbf{\bibinfo{volume}{50}},
  \bibinfo{pages}{454} (\bibinfo{year}{1983}), ISSN \bibinfo{issn}{0031-9007},
  \urlprefix\url{http://link.aps.org/doi/10.1103/PhysRevLett.50.454}.

\bibitem[{\citenamefont{Sokolowski-Tinten and von~der
  Linde}(2000)}]{Sokolowski-Tinten2000}
\bibinfo{author}{\bibfnamefont{K.}~\bibnamefont{Sokolowski-Tinten}}
  \bibnamefont{and} \bibinfo{author}{\bibfnamefont{D.}~\bibnamefont{von~der
  Linde}}, \bibinfo{journal}{Physical Review B} \textbf{\bibinfo{volume}{61}},
  \bibinfo{pages}{2643} (\bibinfo{year}{2000}), ISSN \bibinfo{issn}{0163-1829},
  \urlprefix\url{http://link.aps.org/doi/10.1103/PhysRevB.61.2643}.

\bibitem[{\citenamefont{Rousse et~al.}(2001)\citenamefont{Rousse, Rischel,
  Fourmaux, Uschmann, Sebban, Grillon, Balcou, F\"{o}rster, Geindre, Audebert
  et~al.}}]{Rousse2001}
\bibinfo{author}{\bibfnamefont{A.}~\bibnamefont{Rousse}},
  \bibinfo{author}{\bibfnamefont{C.}~\bibnamefont{Rischel}},
  \bibinfo{author}{\bibfnamefont{S.}~\bibnamefont{Fourmaux}},
  \bibinfo{author}{\bibfnamefont{I.}~\bibnamefont{Uschmann}},
  \bibinfo{author}{\bibfnamefont{S.}~\bibnamefont{Sebban}},
  \bibinfo{author}{\bibfnamefont{G.}~\bibnamefont{Grillon}},
  \bibinfo{author}{\bibfnamefont{P.}~\bibnamefont{Balcou}},
  \bibinfo{author}{\bibfnamefont{E.}~\bibnamefont{F\"{o}rster}},
  \bibinfo{author}{\bibfnamefont{J.~P.} \bibnamefont{Geindre}},
  \bibinfo{author}{\bibfnamefont{P.}~\bibnamefont{Audebert}},
  \bibnamefont{et~al.}, \bibinfo{journal}{Nature}
  \textbf{\bibinfo{volume}{410}}, \bibinfo{pages}{65} (\bibinfo{year}{2001}),
  ISSN \bibinfo{issn}{0028-0836},
  \urlprefix\url{http://dx.doi.org/10.1038/35065045}.

\bibitem[{\citenamefont{Harb et~al.}(2006)\citenamefont{Harb, Ernstorfer,
  Dartigalongue, Hebeisen, Jordan, and Miller}}]{Harb2006}
\bibinfo{author}{\bibfnamefont{M.}~\bibnamefont{Harb}},
  \bibinfo{author}{\bibfnamefont{R.}~\bibnamefont{Ernstorfer}},
  \bibinfo{author}{\bibfnamefont{T.}~\bibnamefont{Dartigalongue}},
  \bibinfo{author}{\bibfnamefont{C.~T.} \bibnamefont{Hebeisen}},
  \bibinfo{author}{\bibfnamefont{R.~E.} \bibnamefont{Jordan}},
  \bibnamefont{and} \bibinfo{author}{\bibfnamefont{R.~J.~D.}
  \bibnamefont{Miller}}, \bibinfo{journal}{The journal of physical chemistry.
  B} \textbf{\bibinfo{volume}{110}}, \bibinfo{pages}{25308}
  (\bibinfo{year}{2006}), ISSN \bibinfo{issn}{1520-6106},
  \urlprefix\url{http://www.ncbi.nlm.nih.gov/pubmed/17165976}.

\bibitem[{\citenamefont{Harb et~al.}(2008)\citenamefont{Harb, Ernstorfer,
  Hebeisen, Sciaini, Peng, Dartigalongue, Eriksson, Lagally, Kruglik, and
  Miller}}]{Harb2008}
\bibinfo{author}{\bibfnamefont{M.}~\bibnamefont{Harb}},
  \bibinfo{author}{\bibfnamefont{R.}~\bibnamefont{Ernstorfer}},
  \bibinfo{author}{\bibfnamefont{C.}~\bibnamefont{Hebeisen}},
  \bibinfo{author}{\bibfnamefont{G.}~\bibnamefont{Sciaini}},
  \bibinfo{author}{\bibfnamefont{W.}~\bibnamefont{Peng}},
  \bibinfo{author}{\bibfnamefont{T.}~\bibnamefont{Dartigalongue}},
  \bibinfo{author}{\bibfnamefont{M.}~\bibnamefont{Eriksson}},
  \bibinfo{author}{\bibfnamefont{M.}~\bibnamefont{Lagally}},
  \bibinfo{author}{\bibfnamefont{S.}~\bibnamefont{Kruglik}}, \bibnamefont{and}
  \bibinfo{author}{\bibfnamefont{R.}~\bibnamefont{Miller}},
  \bibinfo{journal}{Physical Review Letters} \textbf{\bibinfo{volume}{100}},
  \bibinfo{pages}{155504} (\bibinfo{year}{2008}), ISSN
  \bibinfo{issn}{0031-9007},
  \urlprefix\url{http://link.aps.org/doi/10.1103/PhysRevLett.100.155504}.

\bibitem[{\citenamefont{Korfiatis et~al.}(2007)\citenamefont{Korfiatis, Thoma,
  and Vardaxoglou}}]{Korfiatis2007}
\bibinfo{author}{\bibfnamefont{D.~P.} \bibnamefont{Korfiatis}},
  \bibinfo{author}{\bibfnamefont{K.-A.~T.} \bibnamefont{Thoma}},
  \bibnamefont{and} \bibinfo{author}{\bibfnamefont{J.~C.}
  \bibnamefont{Vardaxoglou}}, \bibinfo{journal}{Journal of Physics D: Applied
  Physics} \textbf{\bibinfo{volume}{40}}, \bibinfo{pages}{6803}
  (\bibinfo{year}{2007}), ISSN \bibinfo{issn}{0022-3727},
  \urlprefix\url{http://iopscience.iop.org/0022-3727/40/21/047}.

\bibitem[{\citenamefont{Medvedev and Rethfeld}(2010)}]{Medvedev2010b}
\bibinfo{author}{\bibfnamefont{N.}~\bibnamefont{Medvedev}} \bibnamefont{and}
  \bibinfo{author}{\bibfnamefont{B.}~\bibnamefont{Rethfeld}},
  \bibinfo{journal}{Journal of Applied Physics} \textbf{\bibinfo{volume}{108}},
  \bibinfo{pages}{103112} (\bibinfo{year}{2010}), ISSN
  \bibinfo{issn}{00218979},
  \urlprefix\url{http://link.aip.org/link/JAPIAU/v108/i10/p103112/s1\&Agg=doi}.

\bibitem[{\citenamefont{Gan and Chen}(2011)}]{Gan2011}
\bibinfo{author}{\bibfnamefont{Y.}~\bibnamefont{Gan}} \bibnamefont{and}
  \bibinfo{author}{\bibfnamefont{J.~K.} \bibnamefont{Chen}},
  \bibinfo{journal}{Applied Physics A} \textbf{\bibinfo{volume}{105}},
  \bibinfo{pages}{427} (\bibinfo{year}{2011}), ISSN \bibinfo{issn}{0947-8396},
  \urlprefix\url{http://link.springer.com/10.1007/s00339-011-6573-z}.

\bibitem[{\citenamefont{Gan and Chen}(2013)}]{Gan2013}
\bibinfo{author}{\bibfnamefont{Y.}~\bibnamefont{Gan}} \bibnamefont{and}
  \bibinfo{author}{\bibfnamefont{J.}~\bibnamefont{Chen}},
  \bibinfo{journal}{International Journal of Thermal Sciences}
  \textbf{\bibinfo{volume}{65}}, \bibinfo{pages}{1} (\bibinfo{year}{2013}),
  ISSN \bibinfo{issn}{12900729},
  \urlprefix\url{http://www.sciencedirect.com/science/article/pii/S12900729120%
02670}.

\bibitem[{\citenamefont{Gamaly}(2010)}]{Gamaly2010}
\bibinfo{author}{\bibfnamefont{E.~G.} \bibnamefont{Gamaly}},
  \bibinfo{journal}{Applied Physics A} \textbf{\bibinfo{volume}{101}},
  \bibinfo{pages}{205} (\bibinfo{year}{2010}), ISSN \bibinfo{issn}{0947-8396},
  \urlprefix\url{http://link.springer.com/10.1007/s00339-010-5779-9}.

\bibitem[{\citenamefont{Gamaly and Rode}(2009)}]{Gamaly2009}
\bibinfo{author}{\bibfnamefont{E.~G.} \bibnamefont{Gamaly}} \bibnamefont{and}
  \bibinfo{author}{\bibfnamefont{A.~V.} \bibnamefont{Rode}},
  p.~\bibinfo{pages}{20} (\bibinfo{year}{2009}), \eprint{0910.2150},
  \urlprefix\url{http://arxiv.org/abs/0910.2150}.

\bibitem[{\citenamefont{Ivanov and Zhigilei}(2003)}]{Ivanov2003}
\bibinfo{author}{\bibfnamefont{D.}~\bibnamefont{Ivanov}} \bibnamefont{and}
  \bibinfo{author}{\bibfnamefont{L.}~\bibnamefont{Zhigilei}},
  \bibinfo{journal}{Physical Review B} \textbf{\bibinfo{volume}{68}},
  \bibinfo{pages}{064114} (\bibinfo{year}{2003}), ISSN
  \bibinfo{issn}{0163-1829},
  \urlprefix\url{http://link.aps.org/doi/10.1103/PhysRevB.68.064114}.

\bibitem[{\citenamefont{Lipp et~al.}(2014)\citenamefont{Lipp, Rethfeld, Garcia,
  and Ivanov}}]{Lipp2014a}
\bibinfo{author}{\bibfnamefont{V.~P.} \bibnamefont{Lipp}},
  \bibinfo{author}{\bibfnamefont{B.}~\bibnamefont{Rethfeld}},
  \bibinfo{author}{\bibfnamefont{M.~E.} \bibnamefont{Garcia}},
  \bibnamefont{and} \bibinfo{author}{\bibfnamefont{D.~S.} \bibnamefont{Ivanov}}
  (\bibinfo{year}{2014}), \eprint{1411.4333},
  \urlprefix\url{http://arxiv.org/abs/1411.4333}.

\bibitem[{\citenamefont{Shokeen and Schelling}(2013)}]{Shokeen2013}
\bibinfo{author}{\bibfnamefont{L.}~\bibnamefont{Shokeen}} \bibnamefont{and}
  \bibinfo{author}{\bibfnamefont{P.~K.} \bibnamefont{Schelling}},
  \bibinfo{journal}{Computational Materials Science}
  \textbf{\bibinfo{volume}{67}}, \bibinfo{pages}{316} (\bibinfo{year}{2013}),
  ISSN \bibinfo{issn}{09270256},
  \urlprefix\url{http://www.sciencedirect.com/science/article/pii/S09270256120%
04910}.

\bibitem[{\citenamefont{Medvedev
  et~al.}(2013{\natexlab{a}})\citenamefont{Medvedev, Jeschke, and
  Ziaja}}]{Medvedev2013e}
\bibinfo{author}{\bibfnamefont{N.}~\bibnamefont{Medvedev}},
  \bibinfo{author}{\bibfnamefont{H.~O.} \bibnamefont{Jeschke}},
  \bibnamefont{and} \bibinfo{author}{\bibfnamefont{B.}~\bibnamefont{Ziaja}},
  \bibinfo{journal}{New Journal of Physics} \textbf{\bibinfo{volume}{15}},
  \bibinfo{pages}{015016} (\bibinfo{year}{2013}{\natexlab{a}}), ISSN
  \bibinfo{issn}{1367-2630},
  \urlprefix\url{http://stacks.iop.org/1367-2630/15/i=1/a=015016?key=crossref.%
1a1e1366615a239a23e45c666caddd73}.

\bibitem[{\citenamefont{Medvedev
  et~al.}(2013{\natexlab{b}})\citenamefont{Medvedev, Jeschke, and
  Ziaja}}]{Medvedev2013f}
\bibinfo{author}{\bibfnamefont{N.}~\bibnamefont{Medvedev}},
  \bibinfo{author}{\bibfnamefont{H.~O.} \bibnamefont{Jeschke}},
  \bibnamefont{and} \bibinfo{author}{\bibfnamefont{B.}~\bibnamefont{Ziaja}},
  \bibinfo{journal}{Physical Review B} \textbf{\bibinfo{volume}{88}},
  \bibinfo{pages}{224304} (\bibinfo{year}{2013}{\natexlab{b}}), ISSN
  \bibinfo{issn}{1098-0121},
  \urlprefix\url{http://link.aps.org/doi/10.1103/PhysRevB.88.224304}.

\bibitem[{\citenamefont{Ackermann et~al.}(2007)\citenamefont{Ackermann, Asova,
  Ayvazyan, Azima, Baboi, B\"{a}hr, Balandin, Beutner, Brandt, Bolzmann
  et~al.}}]{Ackermann2007}
\bibinfo{author}{\bibfnamefont{W.}~\bibnamefont{Ackermann}},
  \bibinfo{author}{\bibfnamefont{G.}~\bibnamefont{Asova}},
  \bibinfo{author}{\bibfnamefont{V.}~\bibnamefont{Ayvazyan}},
  \bibinfo{author}{\bibfnamefont{A.}~\bibnamefont{Azima}},
  \bibinfo{author}{\bibfnamefont{N.}~\bibnamefont{Baboi}},
  \bibinfo{author}{\bibfnamefont{J.}~\bibnamefont{B\"{a}hr}},
  \bibinfo{author}{\bibfnamefont{V.}~\bibnamefont{Balandin}},
  \bibinfo{author}{\bibfnamefont{B.}~\bibnamefont{Beutner}},
  \bibinfo{author}{\bibfnamefont{A.}~\bibnamefont{Brandt}},
  \bibinfo{author}{\bibfnamefont{A.}~\bibnamefont{Bolzmann}},
  \bibnamefont{et~al.}, \bibinfo{journal}{Nature Photonics}
  \textbf{\bibinfo{volume}{1}}, \bibinfo{pages}{336} (\bibinfo{year}{2007}),
  ISSN \bibinfo{issn}{1749-4885},
  \urlprefix\url{http://dx.doi.org/10.1038/nphoton.2007.76}.

\bibitem[{\citenamefont{Emma et~al.}(2010)\citenamefont{Emma, Akre, Arthur,
  Bionta, Bostedt, Bozek, Brachmann, Bucksbaum, Coffee, Decker
  et~al.}}]{Emma2010}
\bibinfo{author}{\bibfnamefont{P.}~\bibnamefont{Emma}},
  \bibinfo{author}{\bibfnamefont{R.}~\bibnamefont{Akre}},
  \bibinfo{author}{\bibfnamefont{J.}~\bibnamefont{Arthur}},
  \bibinfo{author}{\bibfnamefont{R.}~\bibnamefont{Bionta}},
  \bibinfo{author}{\bibfnamefont{C.}~\bibnamefont{Bostedt}},
  \bibinfo{author}{\bibfnamefont{J.}~\bibnamefont{Bozek}},
  \bibinfo{author}{\bibfnamefont{A.}~\bibnamefont{Brachmann}},
  \bibinfo{author}{\bibfnamefont{P.}~\bibnamefont{Bucksbaum}},
  \bibinfo{author}{\bibfnamefont{R.}~\bibnamefont{Coffee}},
  \bibinfo{author}{\bibfnamefont{F.-J.} \bibnamefont{Decker}},
  \bibnamefont{et~al.}, \bibinfo{journal}{Nature Photonics}
  \textbf{\bibinfo{volume}{4}}, \bibinfo{pages}{641} (\bibinfo{year}{2010}),
  ISSN \bibinfo{issn}{1749-4885},
  \urlprefix\url{http://dx.doi.org/10.1038/nphoton.2010.176}.

\bibitem[{\citenamefont{Pile}(2011)}]{Pile2011}
\bibinfo{author}{\bibfnamefont{D.}~\bibnamefont{Pile}},
  \bibinfo{journal}{Nature Photonics} \textbf{\bibinfo{volume}{5}},
  \bibinfo{pages}{456} (\bibinfo{year}{2011}), ISSN \bibinfo{issn}{1749-4885},
  \urlprefix\url{http://www.nature.com/nphoton/journal/v5/n8/pdf/nphoton.2011.%
178.pdf?WT.ec\_id=NPHOTON-201108}.

\bibitem[{\citenamefont{Allaria et~al.}(2012)\citenamefont{Allaria, Appio,
  Badano, Barletta, Bassanese, Biedron, Borga, Busetto, Castronovo, Cinquegrana
  et~al.}}]{allaria2012}
\bibinfo{author}{\bibfnamefont{E.}~\bibnamefont{Allaria}},
  \bibinfo{author}{\bibfnamefont{R.}~\bibnamefont{Appio}},
  \bibinfo{author}{\bibfnamefont{L.}~\bibnamefont{Badano}},
  \bibinfo{author}{\bibfnamefont{W.}~\bibnamefont{Barletta}},
  \bibinfo{author}{\bibfnamefont{S.}~\bibnamefont{Bassanese}},
  \bibinfo{author}{\bibfnamefont{S.}~\bibnamefont{Biedron}},
  \bibinfo{author}{\bibfnamefont{A.}~\bibnamefont{Borga}},
  \bibinfo{author}{\bibfnamefont{E.}~\bibnamefont{Busetto}},
  \bibinfo{author}{\bibfnamefont{D.}~\bibnamefont{Castronovo}},
  \bibinfo{author}{\bibfnamefont{P.}~\bibnamefont{Cinquegrana}},
  \bibnamefont{et~al.}, \bibinfo{journal}{Nature Photonics}
  \textbf{\bibinfo{volume}{6}}, \bibinfo{pages}{699} (\bibinfo{year}{2012}),
  ISSN \bibinfo{issn}{1749-4885},
  \urlprefix\url{http://dx.doi.org/10.1038/nphoton.2012.233}.

\bibitem[{\citenamefont{Medvedev et~al.}(2011)\citenamefont{Medvedev, Zastrau,
  F\"{o}rster, Gericke, and Rethfeld}}]{Medvedev2011a}
\bibinfo{author}{\bibfnamefont{N.}~\bibnamefont{Medvedev}},
  \bibinfo{author}{\bibfnamefont{U.}~\bibnamefont{Zastrau}},
  \bibinfo{author}{\bibfnamefont{E.}~\bibnamefont{F\"{o}rster}},
  \bibinfo{author}{\bibfnamefont{D.~O.} \bibnamefont{Gericke}},
  \bibnamefont{and} \bibinfo{author}{\bibfnamefont{B.}~\bibnamefont{Rethfeld}},
  \bibinfo{journal}{Physical Review Letters} \textbf{\bibinfo{volume}{107}},
  \bibinfo{pages}{165003} (\bibinfo{year}{2011}), ISSN
  \bibinfo{issn}{0031-9007},
  \urlprefix\url{http://link.aps.org/doi/10.1103/PhysRevLett.107.165003}.

\bibitem[{\citenamefont{Ziaja and Medvedev}(2012)}]{Ziaja2012}
\bibinfo{author}{\bibfnamefont{B.}~\bibnamefont{Ziaja}} \bibnamefont{and}
  \bibinfo{author}{\bibfnamefont{N.}~\bibnamefont{Medvedev}},
  \bibinfo{journal}{High Energy Density Physics} \textbf{\bibinfo{volume}{8}},
  \bibinfo{pages}{18} (\bibinfo{year}{2012}), ISSN \bibinfo{issn}{15741818},
  \urlprefix\url{http://linkinghub.elsevier.com/retrieve/pii/S1574181811001005%
}.

\bibitem[{\citenamefont{Rethfeld et~al.}(2014)\citenamefont{Rethfeld,
  R\"{a}mer, Brouwer, Medvedev, and Osmani}}]{Rethfeld2014}
\bibinfo{author}{\bibfnamefont{B.}~\bibnamefont{Rethfeld}},
  \bibinfo{author}{\bibfnamefont{A.}~\bibnamefont{R\"{a}mer}},
  \bibinfo{author}{\bibfnamefont{N.}~\bibnamefont{Brouwer}},
  \bibinfo{author}{\bibfnamefont{N.}~\bibnamefont{Medvedev}}, \bibnamefont{and}
  \bibinfo{author}{\bibfnamefont{O.}~\bibnamefont{Osmani}},
  \bibinfo{journal}{Nuclear Inst. and Methods in Physics Research, B}
  \textbf{\bibinfo{volume}{327}}, \bibinfo{pages}{78} (\bibinfo{year}{2014}),
  ISSN \bibinfo{issn}{0168-583X},
  \urlprefix\url{http://dx.doi.org/10.1016/j.nimb.2013.10.087}.

\bibitem[{\citenamefont{Keski-Rahkonen and Krause}(1974)}]{Keski-Rahkonen1974}
\bibinfo{author}{\bibfnamefont{O.}~\bibnamefont{Keski-Rahkonen}}
  \bibnamefont{and} \bibinfo{author}{\bibfnamefont{M.~O.}
  \bibnamefont{Krause}}, \bibinfo{journal}{Atomic Data and Nuclear Data Tables}
  \textbf{\bibinfo{volume}{14}}, \bibinfo{pages}{139} (\bibinfo{year}{1974}),
  ISSN \bibinfo{issn}{0092640X},
  \urlprefix\url{http://dx.doi.org/10.1016/S0092-640X(74)80020-3}.

\bibitem[{\citenamefont{Medvedev and Rethfeld}(2009)}]{Medvedev2009a}
\bibinfo{author}{\bibfnamefont{N.}~\bibnamefont{Medvedev}} \bibnamefont{and}
  \bibinfo{author}{\bibfnamefont{B.}~\bibnamefont{Rethfeld}},
  \bibinfo{journal}{EPL (Europhysics Letters)} \textbf{\bibinfo{volume}{88}},
  \bibinfo{pages}{55001} (\bibinfo{year}{2009}), ISSN
  \bibinfo{issn}{0295-5075},
  \urlprefix\url{http://stacks.iop.org/0295-5075/88/i=5/a=55001?key=crossref.9%
846c6aa960f4e364fbad2e912f22e10}.

\bibitem[{\citenamefont{Ziaja et~al.}(2005)\citenamefont{Ziaja, London, and
  Hajdu}}]{Ziaja2005}
\bibinfo{author}{\bibfnamefont{B.}~\bibnamefont{Ziaja}},
  \bibinfo{author}{\bibfnamefont{R.~A.} \bibnamefont{London}},
  \bibnamefont{and} \bibinfo{author}{\bibfnamefont{J.}~\bibnamefont{Hajdu}},
  \bibinfo{journal}{Journal of Applied Physics} \textbf{\bibinfo{volume}{97}},
  \bibinfo{pages}{064905} (\bibinfo{year}{2005}), ISSN
  \bibinfo{issn}{00218979},
  \urlprefix\url{http://link.aip.org/link/?JAPIAU/97/064905/1}.

\bibitem[{\citenamefont{Lorazo et~al.}(2006)\citenamefont{Lorazo, Lewis, and
  Meunier}}]{Lorazo2006}
\bibinfo{author}{\bibfnamefont{P.}~\bibnamefont{Lorazo}},
  \bibinfo{author}{\bibfnamefont{L.}~\bibnamefont{Lewis}}, \bibnamefont{and}
  \bibinfo{author}{\bibfnamefont{M.}~\bibnamefont{Meunier}},
  \bibinfo{journal}{Physical Review B} \textbf{\bibinfo{volume}{73}},
  \bibinfo{pages}{134108} (\bibinfo{year}{2006}), ISSN
  \bibinfo{issn}{1098-0121},
  \urlprefix\url{http://link.aps.org/doi/10.1103/PhysRevB.73.134108}.

\bibitem[{\citenamefont{Rethfeld et~al.}(2002)\citenamefont{Rethfeld, Kaiser,
  Vicanek, and Simon}}]{Rethfeld2002}
\bibinfo{author}{\bibfnamefont{B.}~\bibnamefont{Rethfeld}},
  \bibinfo{author}{\bibfnamefont{A.}~\bibnamefont{Kaiser}},
  \bibinfo{author}{\bibfnamefont{M.}~\bibnamefont{Vicanek}}, \bibnamefont{and}
  \bibinfo{author}{\bibfnamefont{G.}~\bibnamefont{Simon}},
  \bibinfo{journal}{Physical Review B} \textbf{\bibinfo{volume}{65}},
  \bibinfo{pages}{214303} (\bibinfo{year}{2002}), ISSN
  \bibinfo{issn}{0163-1829},
  \urlprefix\url{http://link.aps.org/doi/10.1103/PhysRevB.65.214303}.

\bibitem[{\citenamefont{Jeschke et~al.}(2001)\citenamefont{Jeschke, Garcia, and
  Bennemann}}]{Jeschke2001}
\bibinfo{author}{\bibfnamefont{H.}~\bibnamefont{Jeschke}},
  \bibinfo{author}{\bibfnamefont{M.}~\bibnamefont{Garcia}}, \bibnamefont{and}
  \bibinfo{author}{\bibfnamefont{K.}~\bibnamefont{Bennemann}},
  \bibinfo{journal}{Physical Review Letters} \textbf{\bibinfo{volume}{87}},
  \bibinfo{pages}{015003} (\bibinfo{year}{2001}), ISSN
  \bibinfo{issn}{0031-9007},
  \urlprefix\url{http://link.aps.org/doi/10.1103/PhysRevLett.87.015003}.

\bibitem[{\citenamefont{Jeschke et~al.}(2002)\citenamefont{Jeschke, Garcia,
  Lenzner, Bonse, Kr\"{u}ger, and Kautek}}]{Jeschke2002}
\bibinfo{author}{\bibfnamefont{H.~O.} \bibnamefont{Jeschke}},
  \bibinfo{author}{\bibfnamefont{M.~E.} \bibnamefont{Garcia}},
  \bibinfo{author}{\bibfnamefont{M.}~\bibnamefont{Lenzner}},
  \bibinfo{author}{\bibfnamefont{J.}~\bibnamefont{Bonse}},
  \bibinfo{author}{\bibfnamefont{J.}~\bibnamefont{Kr\"{u}ger}},
  \bibnamefont{and} \bibinfo{author}{\bibfnamefont{W.}~\bibnamefont{Kautek}},
  \bibinfo{journal}{Applied Surface Science}
  \textbf{\bibinfo{volume}{197-198}}, \bibinfo{pages}{839}
  (\bibinfo{year}{2002}), ISSN \bibinfo{issn}{01694332},
  \urlprefix\url{http://www.sciencedirect.com/science/article/pii/S01694332020%
04580}.

\bibitem[{\citenamefont{Li et~al.}(2013)\citenamefont{Li, Madjet, and
  Vendrell}}]{Li2013a}
\bibinfo{author}{\bibfnamefont{Z.}~\bibnamefont{Li}},
  \bibinfo{author}{\bibfnamefont{M.~E.-A.} \bibnamefont{Madjet}},
  \bibnamefont{and} \bibinfo{author}{\bibfnamefont{O.}~\bibnamefont{Vendrell}},
  \bibinfo{journal}{The Journal of Chemical Physics}
  \textbf{\bibinfo{volume}{138}}, \bibinfo{pages}{094313}
  (\bibinfo{year}{2013}), ISSN \bibinfo{issn}{1089-7690},
  \urlprefix\url{http://scitation.aip.org/content/aip/journal/jcp/138/9/10.106%
3/1.4793274}.

\bibitem[{\citenamefont{Tully}(1990)}]{Tully1990}
\bibinfo{author}{\bibfnamefont{J.~C.} \bibnamefont{Tully}},
  \bibinfo{journal}{The Journal of Chemical Physics}
  \textbf{\bibinfo{volume}{93}}, \bibinfo{pages}{1061} (\bibinfo{year}{1990}),
  ISSN \bibinfo{issn}{00219606},
  \urlprefix\url{http://scitation.aip.org/content/aip/journal/jcp/93/2/10.1063%
/1.459170}.

\bibitem[{\citenamefont{Hammes-Schiffer and Tully}(1994)}]{Hammes-Schiffer1994}
\bibinfo{author}{\bibfnamefont{S.}~\bibnamefont{Hammes-Schiffer}}
  \bibnamefont{and} \bibinfo{author}{\bibfnamefont{J.~C.} \bibnamefont{Tully}},
  \bibinfo{journal}{The Journal of Chemical Physics}
  \textbf{\bibinfo{volume}{101}}, \bibinfo{pages}{4657} (\bibinfo{year}{1994}),
  ISSN \bibinfo{issn}{00219606},
  \urlprefix\url{http://scitation.aip.org/content/aip/journal/jcp/101/6/10.106%
3/1.467455}.

\bibitem[{\citenamefont{Medvedev et~al.}(2015)\citenamefont{Medvedev,
  Tkachenko, and Ziaja}}]{Medvedev2015}
\bibinfo{author}{\bibfnamefont{N.}~\bibnamefont{Medvedev}},
  \bibinfo{author}{\bibfnamefont{V.}~\bibnamefont{Tkachenko}},
  \bibnamefont{and} \bibinfo{author}{\bibfnamefont{B.}~\bibnamefont{Ziaja}},
  \bibinfo{journal}{Contributions to Plasma Physics}
  \textbf{\bibinfo{volume}{55}}, \bibinfo{pages}{12} (\bibinfo{year}{2015}),
  ISSN \bibinfo{issn}{08631042},
  \urlprefix\url{http://doi.wiley.com/10.1002/ctpp.201400026}.

\bibitem[{\citenamefont{Jeschke et~al.}(1999)\citenamefont{Jeschke, Garcia, and
  Bennemann}}]{Jeschke1999}
\bibinfo{author}{\bibfnamefont{H.}~\bibnamefont{Jeschke}},
  \bibinfo{author}{\bibfnamefont{M.}~\bibnamefont{Garcia}}, \bibnamefont{and}
  \bibinfo{author}{\bibfnamefont{K.}~\bibnamefont{Bennemann}},
  \bibinfo{journal}{Physical Review B} \textbf{\bibinfo{volume}{60}},
  \bibinfo{pages}{R3701} (\bibinfo{year}{1999}), ISSN
  \bibinfo{issn}{0163-1829},
  \urlprefix\url{http://link.aps.org/doi/10.1103/PhysRevB.60.R3701}.

\bibitem[{\citenamefont{Parrinello and Rahman}(1980)}]{Parrinello1980}
\bibinfo{author}{\bibfnamefont{M.}~\bibnamefont{Parrinello}} \bibnamefont{and}
  \bibinfo{author}{\bibfnamefont{A.}~\bibnamefont{Rahman}},
  \bibinfo{journal}{Physical Review Letters} \textbf{\bibinfo{volume}{45}},
  \bibinfo{pages}{1196} (\bibinfo{year}{1980}), ISSN \bibinfo{issn}{0031-9007},
  \urlprefix\url{http://link.aps.org/doi/10.1103/PhysRevLett.45.1196}.

\bibitem[{\citenamefont{Chapman and Gericke}(2011)}]{Chapman2011}
\bibinfo{author}{\bibfnamefont{D.~A.} \bibnamefont{Chapman}} \bibnamefont{and}
  \bibinfo{author}{\bibfnamefont{D.~O.} \bibnamefont{Gericke}},
  \bibinfo{journal}{Physical Review Letters} \textbf{\bibinfo{volume}{107}},
  \bibinfo{pages}{165004} (\bibinfo{year}{2011}), ISSN
  \bibinfo{issn}{0031-9007},
  \urlprefix\url{http://link.aps.org/doi/10.1103/PhysRevLett.107.165004}.

\bibitem[{\citenamefont{F\"{a}ustlin et~al.}(2010)\citenamefont{F\"{a}ustlin,
  Bornath, D\"{o}ppner, D\"{u}sterer, F\"{o}rster, Fortmann, Glenzer, G\"{o}de,
  Gregori, Irsig et~al.}}]{Faustlin2010}
\bibinfo{author}{\bibfnamefont{R.~R.} \bibnamefont{F\"{a}ustlin}},
  \bibinfo{author}{\bibfnamefont{T.}~\bibnamefont{Bornath}},
  \bibinfo{author}{\bibfnamefont{T.}~\bibnamefont{D\"{o}ppner}},
  \bibinfo{author}{\bibfnamefont{S.}~\bibnamefont{D\"{u}sterer}},
  \bibinfo{author}{\bibfnamefont{E.}~\bibnamefont{F\"{o}rster}},
  \bibinfo{author}{\bibfnamefont{C.}~\bibnamefont{Fortmann}},
  \bibinfo{author}{\bibfnamefont{S.~H.} \bibnamefont{Glenzer}},
  \bibinfo{author}{\bibfnamefont{S.}~\bibnamefont{G\"{o}de}},
  \bibinfo{author}{\bibfnamefont{G.}~\bibnamefont{Gregori}},
  \bibinfo{author}{\bibfnamefont{R.}~\bibnamefont{Irsig}},
  \bibnamefont{et~al.}, \bibinfo{journal}{Physical Review Letters}
  \textbf{\bibinfo{volume}{104}}, \bibinfo{pages}{125002}
  (\bibinfo{year}{2010}), ISSN \bibinfo{issn}{0031-9007},
  \urlprefix\url{http://link.aps.org/doi/10.1103/PhysRevLett.104.125002}.

\bibitem[{\citenamefont{Hau-Riege}(2013)}]{Hau-Riege2013}
\bibinfo{author}{\bibfnamefont{S.~P.} \bibnamefont{Hau-Riege}},
  \bibinfo{journal}{Physical Review E} \textbf{\bibinfo{volume}{87}},
  \bibinfo{pages}{053102} (\bibinfo{year}{2013}), ISSN
  \bibinfo{issn}{1539-3755},
  \urlprefix\url{http://link.aps.org/doi/10.1103/PhysRevE.87.053102}.

\bibitem[{\citenamefont{Rymzhanov et~al.}(2014)\citenamefont{Rymzhanov,
  Medvedev, and Volkov}}]{Rymzhanov2014a}
\bibinfo{author}{\bibfnamefont{R.~A.} \bibnamefont{Rymzhanov}},
  \bibinfo{author}{\bibfnamefont{N.~A.} \bibnamefont{Medvedev}},
  \bibnamefont{and} \bibinfo{author}{\bibfnamefont{A.~E.}
  \bibnamefont{Volkov}}, \bibinfo{journal}{physica status solidi (b)} pp.
  \bibinfo{pages}{n/a--n/a} (\bibinfo{year}{2014}), ISSN
  \bibinfo{issn}{03701972},
  \urlprefix\url{http://doi.wiley.com/10.1002/pssb.201400130}.

\bibitem[{\citenamefont{Osmani et~al.}(2011)\citenamefont{Osmani, Medvedev,
  Schleberger, and Rethfeld}}]{Osmani2011}
\bibinfo{author}{\bibfnamefont{O.}~\bibnamefont{Osmani}},
  \bibinfo{author}{\bibfnamefont{N.}~\bibnamefont{Medvedev}},
  \bibinfo{author}{\bibfnamefont{M.}~\bibnamefont{Schleberger}},
  \bibnamefont{and} \bibinfo{author}{\bibfnamefont{B.}~\bibnamefont{Rethfeld}},
  \bibinfo{journal}{Physical Review B} \textbf{\bibinfo{volume}{84}},
  \bibinfo{pages}{214105} (\bibinfo{year}{2011}), ISSN
  \bibinfo{issn}{1098-0121},
  \urlprefix\url{http://link.aps.org/doi/10.1103/PhysRevB.84.214105}.

\bibitem[{\citenamefont{Ridgway et~al.}(2013)\citenamefont{Ridgway, Bierschenk,
  Giulian, Afra, Rodriguez, Araujo, Byrne, Kirby, Pakarinen, Djurabekova
  et~al.}}]{Ridgway2013}
\bibinfo{author}{\bibfnamefont{M.~C.} \bibnamefont{Ridgway}},
  \bibinfo{author}{\bibfnamefont{T.}~\bibnamefont{Bierschenk}},
  \bibinfo{author}{\bibfnamefont{R.}~\bibnamefont{Giulian}},
  \bibinfo{author}{\bibfnamefont{B.}~\bibnamefont{Afra}},
  \bibinfo{author}{\bibfnamefont{M.~D.} \bibnamefont{Rodriguez}},
  \bibinfo{author}{\bibfnamefont{L.~L.} \bibnamefont{Araujo}},
  \bibinfo{author}{\bibfnamefont{a.~P.} \bibnamefont{Byrne}},
  \bibinfo{author}{\bibfnamefont{N.}~\bibnamefont{Kirby}},
  \bibinfo{author}{\bibfnamefont{O.~H.} \bibnamefont{Pakarinen}},
  \bibinfo{author}{\bibfnamefont{F.}~\bibnamefont{Djurabekova}},
  \bibnamefont{et~al.}, \bibinfo{journal}{Physical Review Letters}
  \textbf{\bibinfo{volume}{110}}, \bibinfo{pages}{245502}
  (\bibinfo{year}{2013}), ISSN \bibinfo{issn}{0031-9007},
  \urlprefix\url{http://link.aps.org/doi/10.1103/PhysRevLett.110.245502}.

\bibitem[{\citenamefont{Rethfeld et~al.}(2010)\citenamefont{Rethfeld, Brenk,
  Medvedev, Krutsch, and Hoffmann}}]{Rethfeld2010}
\bibinfo{author}{\bibfnamefont{B.}~\bibnamefont{Rethfeld}},
  \bibinfo{author}{\bibfnamefont{O.}~\bibnamefont{Brenk}},
  \bibinfo{author}{\bibfnamefont{N.}~\bibnamefont{Medvedev}},
  \bibinfo{author}{\bibfnamefont{H.}~\bibnamefont{Krutsch}}, \bibnamefont{and}
  \bibinfo{author}{\bibfnamefont{D.~H.~H.} \bibnamefont{Hoffmann}},
  \bibinfo{journal}{Applied Physics A} \textbf{\bibinfo{volume}{101}},
  \bibinfo{pages}{19} (\bibinfo{year}{2010}), ISSN \bibinfo{issn}{0947-8396},
  \urlprefix\url{http://link.springer.com/10.1007/s00339-010-5780-3}.

\bibitem[{\citenamefont{Medvedev
  et~al.}(2013{\natexlab{c}})\citenamefont{Medvedev, Jeschke, and
  Ziaja}}]{Medvedev2013}
\bibinfo{author}{\bibfnamefont{N.~A.} \bibnamefont{Medvedev}},
  \bibinfo{author}{\bibfnamefont{H.~O.} \bibnamefont{Jeschke}},
  \bibnamefont{and} \bibinfo{author}{\bibfnamefont{B.}~\bibnamefont{Ziaja}},
  \bibinfo{journal}{SPIE Proc.} \textbf{\bibinfo{volume}{8777}},
  \bibinfo{pages}{877709} (\bibinfo{year}{2013}{\natexlab{c}}).

\bibitem[{\citenamefont{{Harald O. Jeschke}}(2000)}]{HaraldO.Jeschke2000}
\bibinfo{author}{\bibnamefont{{Harald O. Jeschke}}}, Ph.D. thesis,
  \bibinfo{school}{Technical University of Berlin} (\bibinfo{year}{2000}),
  \urlprefix\url{http://www.physics.rutgers.edu/~jeschke/phd.html}.

\bibitem[{\citenamefont{Mueller and Rethfeld}(2013)}]{Mueller2013}
\bibinfo{author}{\bibfnamefont{B.~Y.} \bibnamefont{Mueller}} \bibnamefont{and}
  \bibinfo{author}{\bibfnamefont{B.}~\bibnamefont{Rethfeld}},
  \bibinfo{journal}{Physical Review B} \textbf{\bibinfo{volume}{87}},
  \bibinfo{pages}{035139} (\bibinfo{year}{2013}), ISSN
  \bibinfo{issn}{1098-0121},
  \urlprefix\url{http://link.aps.org/doi/10.1103/PhysRevB.87.035139}.

\bibitem[{\citenamefont{Rapaport}(2004)}]{Rapaport2004}
\bibinfo{author}{\bibfnamefont{D.~C.} \bibnamefont{Rapaport}},
  \emph{\bibinfo{title}{{The Art of Molecular Dynamics Simulation}}}
  (\bibinfo{publisher}{Cambridge University Press}, \bibinfo{year}{2004}), ISBN
  \bibinfo{isbn}{0521825687},
  \urlprefix\url{http://books.google.com/books?id=iqDJ2hjqBMEC\&pgis=1}.

\bibitem[{\citenamefont{Chen et~al.}(2005)\citenamefont{Chen, Tzou, and
  Beraun}}]{Chen2005}
\bibinfo{author}{\bibfnamefont{J.}~\bibnamefont{Chen}},
  \bibinfo{author}{\bibfnamefont{D.}~\bibnamefont{Tzou}}, \bibnamefont{and}
  \bibinfo{author}{\bibfnamefont{J.}~\bibnamefont{Beraun}},
  \bibinfo{journal}{International Journal of Heat and Mass Transfer}
  \textbf{\bibinfo{volume}{48}}, \bibinfo{pages}{501} (\bibinfo{year}{2005}),
  ISSN \bibinfo{issn}{00179310},
  \urlprefix\url{http://dx.doi.org/10.1016/j.ijheatmasstransfer.2004.09.015}.

\bibitem[{\citenamefont{Beye et~al.}(2013)\citenamefont{Beye, Wernet,
  Sch\"{u}\ss~ler Langeheine, and F\"{o}hlisch}}]{Beye2013}
\bibinfo{author}{\bibfnamefont{M.}~\bibnamefont{Beye}},
  \bibinfo{author}{\bibfnamefont{P.}~\bibnamefont{Wernet}},
  \bibinfo{author}{\bibfnamefont{C.}~\bibnamefont{Sch\"{u}\ss~ler Langeheine}},
  \bibnamefont{and}
  \bibinfo{author}{\bibfnamefont{A.}~\bibnamefont{F\"{o}hlisch}},
  \bibinfo{journal}{Journal of Electron Spectroscopy and Related Phenomena}
  \textbf{\bibinfo{volume}{188}}, \bibinfo{pages}{172} (\bibinfo{year}{2013}),
  ISSN \bibinfo{issn}{03682048},
  \urlprefix\url{http://www.sciencedirect.com/science/article/pii/S03682048130%
00789}.

\bibitem[{\citenamefont{Sokolowski-Tinten
  et~al.}(1995)\citenamefont{Sokolowski-Tinten, Bialkowski, and von~der
  Linde}}]{Sokolowski-Tinten1995}
\bibinfo{author}{\bibfnamefont{K.}~\bibnamefont{Sokolowski-Tinten}},
  \bibinfo{author}{\bibfnamefont{J.}~\bibnamefont{Bialkowski}},
  \bibnamefont{and} \bibinfo{author}{\bibfnamefont{D.}~\bibnamefont{von~der
  Linde}}, \bibinfo{journal}{Physical Review B} \textbf{\bibinfo{volume}{51}},
  \bibinfo{pages}{14186} (\bibinfo{year}{1995}).

\bibitem[{\citenamefont{Palik}(1985)}]{Palik1985}
\bibinfo{author}{\bibfnamefont{E.~D.} \bibnamefont{Palik}},
  \emph{\bibinfo{title}{{Handbook of Optical Constants of Solids}}},
  vol.~\bibinfo{volume}{1} of \emph{\bibinfo{series}{Academic Press handbook
  series}} (\bibinfo{publisher}{Academic Press}, \bibinfo{year}{1985}), ISBN
  \bibinfo{isbn}{0125444206},
  \urlprefix\url{http://www.sciencedirect.com/science/book/9780125444156}.

\bibitem[{\citenamefont{Henke et~al.}(1993)\citenamefont{Henke, Gullikson, and
  Davis}}]{Henke1993}
\bibinfo{author}{\bibfnamefont{B.}~\bibnamefont{Henke}},
  \bibinfo{author}{\bibfnamefont{E.}~\bibnamefont{Gullikson}},
  \bibnamefont{and} \bibinfo{author}{\bibfnamefont{J.}~\bibnamefont{Davis}},
  \bibinfo{journal}{Atomic Data and Nuclear Data Tables}
  \textbf{\bibinfo{volume}{54}}, \bibinfo{pages}{181} (\bibinfo{year}{1993}),
  ISSN \bibinfo{issn}{0092640X},
  \urlprefix\url{http://dx.doi.org/10.1006/adnd.1993.1013}.

\bibitem[{\citenamefont{Cullen et~al.}(1997)\citenamefont{Cullen, Hubbell, and
  Kissel}}]{Cullen1997}
\bibinfo{author}{\bibfnamefont{D.~E.} \bibnamefont{Cullen}},
  \bibinfo{author}{\bibfnamefont{J.~H.} \bibnamefont{Hubbell}},
  \bibnamefont{and} \bibinfo{author}{\bibfnamefont{L.}~\bibnamefont{Kissel}},
  \emph{\bibinfo{title}{{EPDL97: the Evaluated Photon Data Library, '97
  version.}}} (\bibinfo{publisher}{Lawrence Livermore National Laboratory,
  UCRL--50400}, \bibinfo{address}{Livermore, CA}, \bibinfo{year}{1997}),
  \bibinfo{edition}{vol. 6, re} ed.

\bibitem[{\citenamefont{Beye et~al.}(2010)\citenamefont{Beye, Sorgenfrei,
  Schlotter, Wurth, and F\"{o}hlisch}}]{Beye2010}
\bibinfo{author}{\bibfnamefont{M.}~\bibnamefont{Beye}},
  \bibinfo{author}{\bibfnamefont{F.}~\bibnamefont{Sorgenfrei}},
  \bibinfo{author}{\bibfnamefont{W.~F.} \bibnamefont{Schlotter}},
  \bibinfo{author}{\bibfnamefont{W.}~\bibnamefont{Wurth}}, \bibnamefont{and}
  \bibinfo{author}{\bibfnamefont{A.}~\bibnamefont{F\"{o}hlisch}},
  \bibinfo{journal}{Proceedings of the National Academy of Sciences of the
  United States of America} \textbf{\bibinfo{volume}{107}},
  \bibinfo{pages}{16772} (\bibinfo{year}{2010}), ISSN
  \bibinfo{issn}{1091-6490},
  \urlprefix\url{http://www.pnas.org/content/107/39/16772.full}.

\bibitem[{\citenamefont{Plasser et~al.}(2012)\citenamefont{Plasser, Granucci,
  Pittner, Barbatti, Persico, and Lischka}}]{Plasser2012}
\bibinfo{author}{\bibfnamefont{F.}~\bibnamefont{Plasser}},
  \bibinfo{author}{\bibfnamefont{G.}~\bibnamefont{Granucci}},
  \bibinfo{author}{\bibfnamefont{J.}~\bibnamefont{Pittner}},
  \bibinfo{author}{\bibfnamefont{M.}~\bibnamefont{Barbatti}},
  \bibinfo{author}{\bibfnamefont{M.}~\bibnamefont{Persico}}, \bibnamefont{and}
  \bibinfo{author}{\bibfnamefont{H.}~\bibnamefont{Lischka}},
  \bibinfo{journal}{The Journal of chemical physics}
  \textbf{\bibinfo{volume}{137}}, \bibinfo{pages}{22A514}
  (\bibinfo{year}{2012}), ISSN \bibinfo{issn}{1089-7690},
  \urlprefix\url{http://scitation.aip.org/content/aip/journal/jcp/137/22/10.10%
63/1.4738960}.

\end{thebibliography}
\end{document}